\documentclass[aps,pre,reprint,groupedaddress,longbibliography]{revtex4-1}

\usepackage[english]{babel}
\usepackage[utf8]{inputenc}
\usepackage{hyperref}

\usepackage{amsmath,amssymb,mathtools,mathdots,bm,bbm,upgreek}

\usepackage[usenames,dvipsnames]{xcolor}
\usepackage{graphicx}
\usepackage[caption=false]{subfig}
\usepackage{floatrow}
\usepackage[shortlabels,inline]{enumitem}

\DeclarePairedDelimiterX{\braket}[2]{\langle}{\rangle}{#1\,\delimsize\mvert\,\mathopen{}#2}
\DeclarePairedDelimiterX{\ketbra}[2]{\vert}{\vert}{#1\,\delimsize\rangle\!\delimsize\langle\,\mathopen{}#2}
\DeclarePairedDelimiterX{\comm}[2]{[}{]}{#1, #2}
\DeclarePairedDelimiterX{\acomm}[2]{\{}{\}}{#1, #2}

\newcommand{\dd}{\textrm d}
\newcommand{\id}{\;\dd}
\newcommand{\dt}{\dd t}
\newcommand{\e}{\textrm e}
\renewcommand{\vec}[1]{\bm{#1}}
\newcommand{\vectwo}[2]{\ensuremath \begin{pmatrix}#1\\#2\end{pmatrix}}

\DeclareMathOperator{\tr}{tr}

\renewcommand{\log}{\ln}

\newcommand{\txtc}{\text{c}}
\newcommand{\txth}{\text{h}}
\newcommand{\txtmin}{\text{min}}
\newcommand{\txtmax}{\text{max}}
\newcommand{\eq}{\text{eq}}
\newcommand{\switch}[1]{{#1}^\prime}
\newcommand{\period}{\tau}

\newcommand{\fixedOutput}{\mathcal Q_\txtc^\ast}
\newcommand{\durationFunctional}{\mathcal T}
\newcommand{\outputX}[1]{{#1}^{\text{p}}}
\newcommand{\effX}[1]{{#1}^{\upeta}}
\newcommand{\Xwork}[1]{{#1}_{\text{w}}}
\newcommand{\Xreset}[1]{{#1}_{\text{r}}}
\newcommand{\depart}{1}

\begin{document}

\title{Two-Stroke Optimization Scheme for Mesoscopic Refrigerators}

\author{Paul Menczel}

\author{Tuomas Pyhäranta}
\author{Christian Flindt}
\author{Kay Brandner}
\affiliation{Department of Applied Physics, Aalto University, 00076 Aalto, Finland}



\begin{abstract}
Refrigerators use a thermodynamic cycle to move thermal energy from a cold reservoir to a hot one.
Implementing this operation principle with mesoscopic components has recently emerged as a promising strategy to control heat currents in micro and nano systems for quantum technological applications.
Here, we combine concepts from stochastic and quantum thermodynamics with advanced methods of optimal control theory to develop a universal optimization scheme for such small-scale refrigerators.
Covering both the classical and the quantum regime, our theoretical framework provides a rigorous procedure to determine the periodic driving protocols that maximize either cooling power or efficiency.
As a main technical tool, we decompose the cooling cycle into two strokes, which can be optimized one by one.
In the regimes of slow or fast driving, we show how this procedure can be simplified significantly by invoking suitable approximations.
To demonstrate the practical viability of our scheme, we determine the exact optimal driving protocols for a quantum microcooler, which can be realized experimentally with current technology.
Our work provides a powerful tool to develop optimal design strategies for engineered cooling devices and it creates a versatile framework for theoretical investigations exploring the fundamental performance limits of mesoscopic thermal machines.
\end{abstract}


\maketitle


\section{Introduction} \label{sec:introduction}

With the rapid advance of quantum technologies during the last decade, the search for new strategies to overcome the challenges of thermal management at low temperatures and small length scales has become a subject of intense research \cite{GiazottoRevModPhys2006,LindenPhysRevLett2010,MuhonenRepProgPhys2012,CourtoisJLowTempPhys2014,PekolaNatPhys2015}.
Solid-state quantum devices based on, for example, superconducting circuits require operation temperatures in the range of millikelvins, which must currently be upheld with massive and costly cryogenic equipment.
These systems are among the most promising candidates to realize a large-scale quantum computer \cite{DevoretScience2013,BarendsNature2014,KellyNature2015}; 
	they also provide a versatile platform for the design of accurately tunable thermal instruments that can be implemented on chip and thus make it possible to control the heat flow between individual components of complex quantum circuits \cite{GrajcarNatPhys2008,GiazottoNature2012,HoferPhysRevB2016a,PartanenNatPhys2016,FornieriNatNanotechnol2016,TanNatCommun2017,RonzaniNatPhys2018}. 
This technology could significantly simplify the operation of quantum devices by enabling the selective cooling of their functional degrees of freedom.

\begin{figure}
	\centering
	\includegraphics[scale=1]{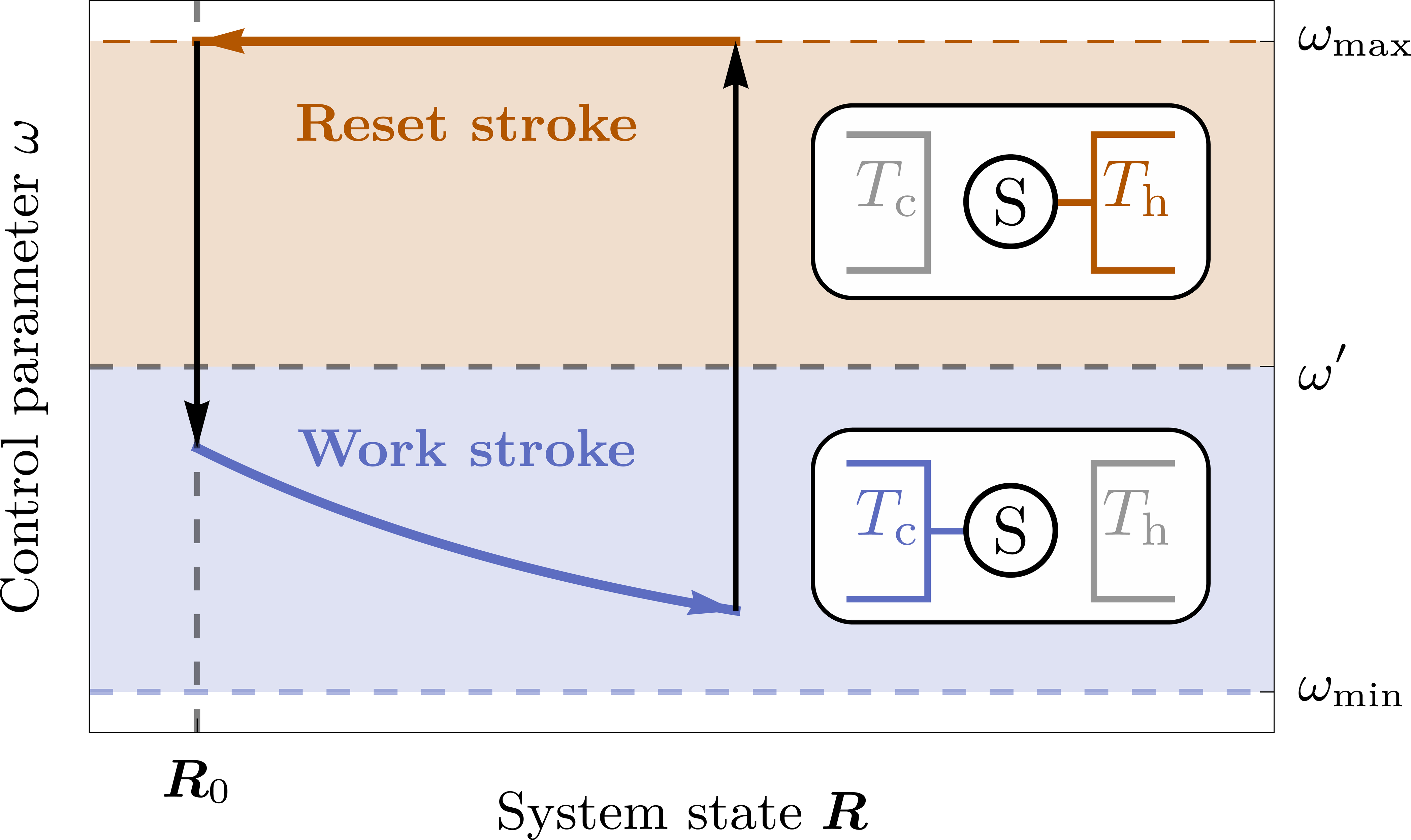}
	\caption{Thermodynamic operation cycle of a two-stroke refrigerator.
		Depending on whether the value of the control parameter $\omega$ is smaller or larger than a given threshold $\switch\omega$, the working system S couples to a reservoir with temperature $T_\txtc$ or to a hotter one with temperature $T_\txth > T_\txtc$.
		The possible values of the control parameter are delimited by $\omega_\txtmin$ and $\omega_\txtmax$.
		In the work stroke, heat is transferred from the cold reservoir to the working system (blue arrow).
		The reset stroke restores the initial state $\vec R_0$, while S is in contact with the hot reservoir (red arrow).
		The two strokes are connected by instantaneous jumps of the control parameter (black arrows).}
	\label{fig:introduction:cycle}
\end{figure}

Mesoscopic refrigerators play a promising role in the development of integrated quantum cooling solutions. 
Mimicking the cyclic operation principle of their macroscopic counterparts, which are used in everyday appliances such as freezers and air conditioners, these devices use periodic driving fields to transfer heat from a cold object to a hot one \cite{QuanPhysRevLett2006,NiskanenPhysRevB2007,AllahverdyanPhysRevE2010,KosloffPhysRevE2010,FeldmannPhysRevE2012,KolarPhysRevLett2012,IzumidaEPL2013,CampisiNewJPhys2015,UzdinPhysRevX2015,
AbahEPL2016,ProesmansPhysRevX2016,PekolaArXiv181210933Quant-Ph2018}.
Their basic working mechanism can be understood as a two-stroke process. 
In the first stroke, a certain amount of heat is absorbed from the cold body into a working system, which acts as a container for thermal energy. 
The second stroke uses the power input from the external driving field to inject the acquired heat into a hot reservoir and restore the initial state of the working system as illustrated in Fig.~\ref{fig:introduction:cycle}.

The thermodynamic performance of this cycle is crucially determined by the driving protocol that is applied to the working system. 
Finding its optimal shape is vital for practical applications and, at the same time, constitutes a formidable theoretical task. 
In fact, finding optimal strategies to control periodic thermodynamic processes in small-scale systems is a longstanding problem in both stochastic \cite{SchmiedlEPL2007,EspositoEPL2010,AndresenAngewChemIntEd2011,HolubecJStatMech2014,DechantPhysRevLett2015,BrandnerPhysRevX2015,BauerPhysRevE2016,DechantEPL2017} and quantum thermodynamics \cite{UzdinEPL2014,BrandnerPhysRevE2016,KarimiPhysRevB2016,SuriEurPhysJSpecTop2018,CavinaPhysRevA2018,ErdmanArXiv181205089Quant-Ph2018}, which involves three major challenges. 
First, the intricate interdependence between state and control variables that governs the dynamics of mesoscopic devices leads to constraints that can usually not be solved explicitly.
Second, thermodynamic figures of merit such as cooling power are typically unbounded functions of external control parameters.
The optimal protocol is then determined by the boundaries of the admissible parameter space and cannot be found from Euler-Lagrange equations, a situation known as a bang-bang scenario \cite{Kirk2004,BoldtEPL2012,DeffnerJPhysB2014}.
Third, a periodic mode of operation requires that the initial configuration of the device is restored after a given cycle time \cite{KosloffEntropy2017,CavinaPhysRevA2018}.
This constraint effectively renders the optimization problem non-local in time, since any change of the driving protocol during the cycle affects the final state of the working system.

In this article, we show how these problems can be handled in three successive steps forming a universal scheme that makes it possible to maximize both the cooling power and the efficiency of mesoscopic refrigerators.
The key idea of our method is to divide the refrigeration cycle into two strokes, which can be optimized one by one after fixing suitable boundary conditions, see Fig.~\ref{fig:introduction:cycle}.
Dynamical constraints are thereby included through time-dependent Lagrange multipliers and bang-bang type protocols are taken into account systematically by applying Pontryagin's minimum principle \cite{Pontryagin1962,Kirk2004} as we explain in the following.
This two-step procedure effectively fixes the shape of the optimal driving protocol. 
The extracted heat, which initially depends on the entire control protocol, is thus reduced to an ordinary function of time-independent variational parameters, which can be optimized with standard techniques. 

To illustrate our general formalism, we analyze a semiclassical model of a realistic quantum microcooler based on superconducting circuits, which can be implemented with current experimental technology \cite{KarimiPhysRevB2016,RonzaniNatPhys2018}.
This application demonstrates the practical viability of our new scheme.
Moreover, since the optimization of our model can be performed essentially through analytical calculations, it also provides valuable insights into characteristic features of optimal cooling cycles in mesoscopic systems.

The scope of our two-stroke framework is not limited to elementary models that can be treated exactly.
By contrast, owing to its general structure, our scheme can be combined with a variety of established dynamical approximation methods to become an even more powerful theoretical tool.
In this way, a physically transparent picture can also be obtained of complicated optimization problems, for which even numerically exact solutions would be practically out of reach.
In the second part of our paper, we show how such a perturbative approach can be implemented for the limiting regimes of slow and fast driving. 
We round off our work by applying these techniques to determine the optimal working conditions of a superconducting microcooler in the full quantum regime.

Our manuscript is organized as follows.
In Sec.~\ref{sec:general}, we establish our two-stroke optimization scheme, which provides the general basis for this paper.
In Sec.~\ref{sec:fridge}, we use this framework to optimize the performance of a realistic model for a quantum microcooler in the semiclassical regime.
We further develop our general theory in Sec.~\ref{sec:approx} by incorporating two key dynamical approximation methods.
In Sec.~\ref{sec:coh}, we apply these techniques to extend the semiclassical case study of Sec.~\ref{sec:fridge} to the coherent regime.
Finally, we conclude and discuss the new perspectives opened by our work in Sec.~\ref{sec:outlook}.
Appendices~\ref{sec:appendix1} and~\ref{sec:appendix2} contain further technical details of our calculations.

\section{General Scheme} \label{sec:general}

\begin{figure*}
	\centering
	\null\hfill
	\includegraphics[scale=1]{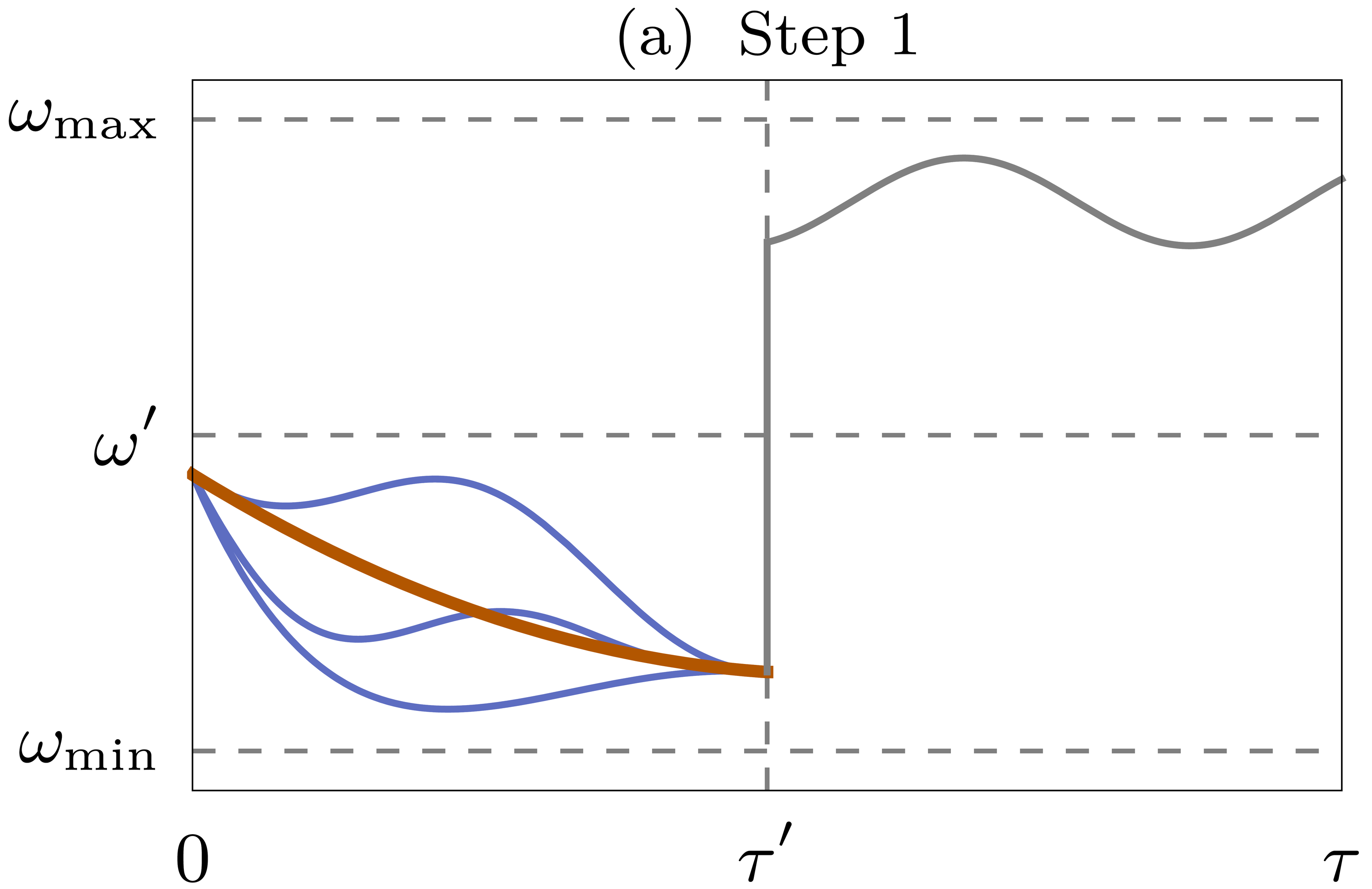}
	\hfill
	\includegraphics[scale=1]{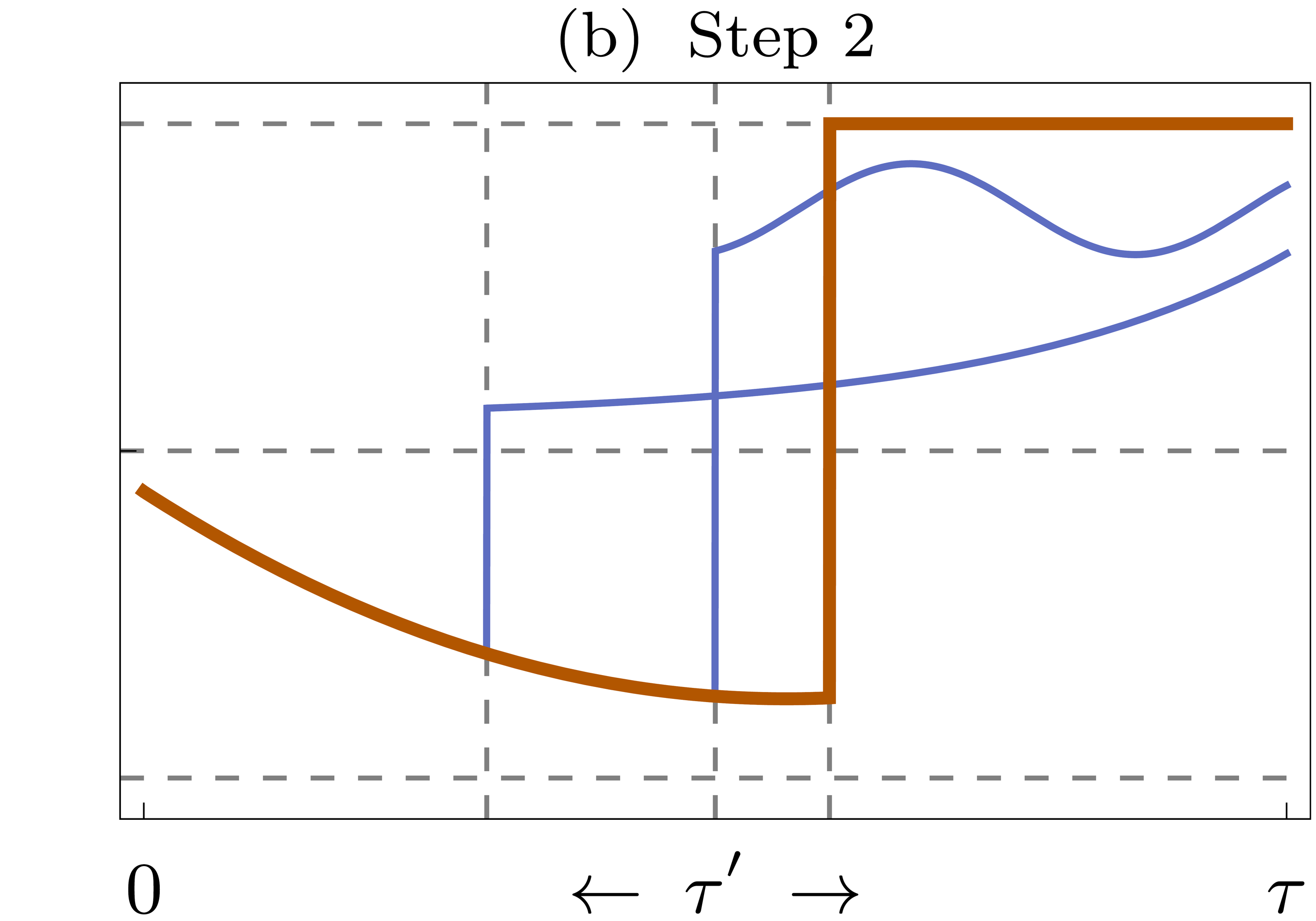}
	\hfill
	\includegraphics[scale=1]{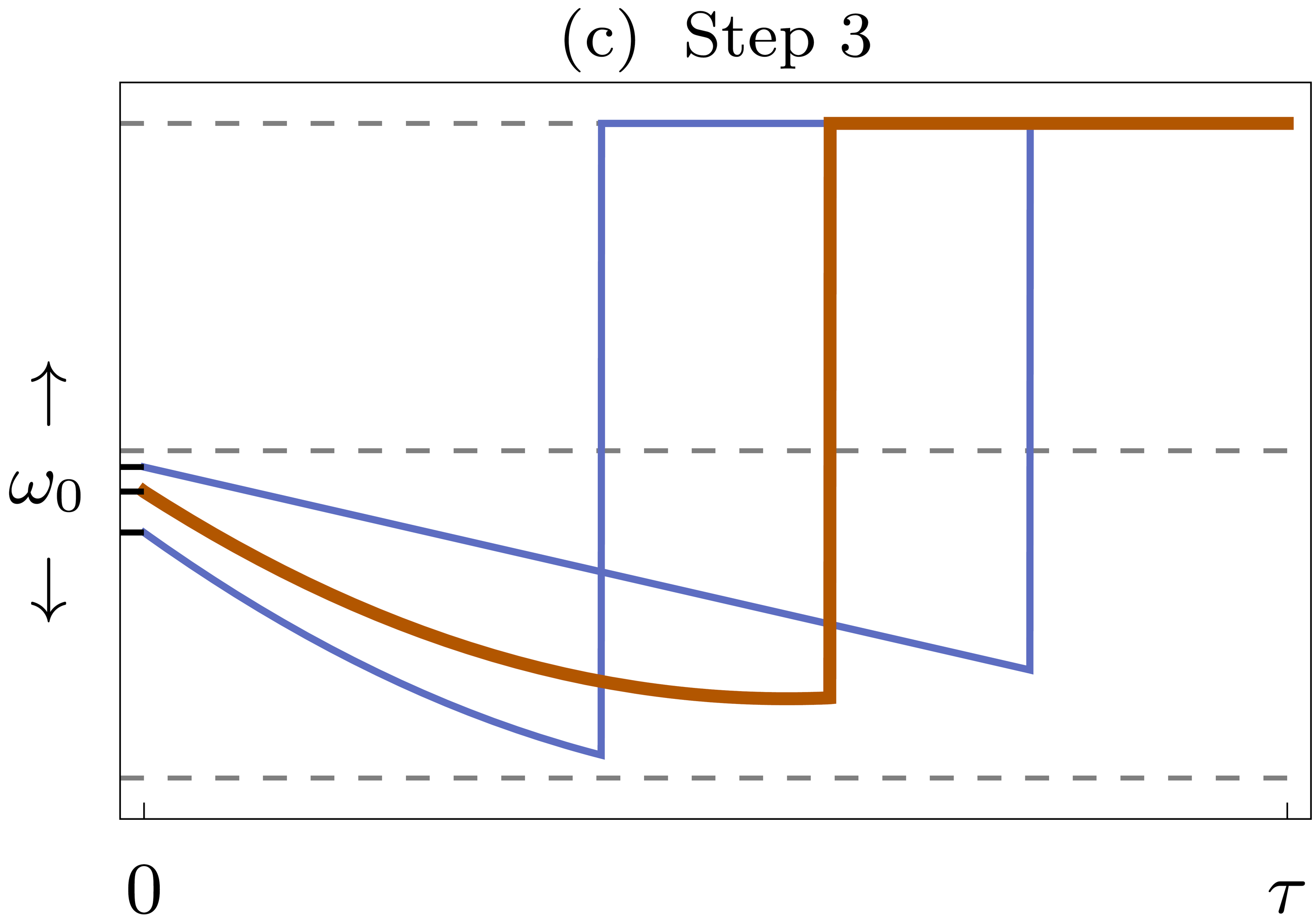}
	\hfill\null
	\caption[this_is_necessary]{Maximizing the cooling power of a two-stroke refrigerator in three steps.
		\begin{enumerate*}[(a)]
		\item In step 1, the optimal work protocol (red line) is determined by variation of the functional \eqref{eq:gen:ext_output} for a fixed switching time $\switch\period$.
			Adding small displacements (blue lines) to the optimal protocol can only reduce the extracted heat.
			The reset stroke protocol (gray line) does not play a role here.
		\item Step 2 optimizes the reset protocol such that the initial state of the system is restored, while keeping the optimal work stroke with given initial conditions fixed.
			The optimal reset stroke (in red) is thereby distinguished by having the latest possible switching time $\switch\period$.
			The blue lines are examples of non-optimal reset stroke protocols with earlier switching times.
		\item The initial values are determined in step 3, which completes the optimal protocol (in red).
			It is here compared to the protocols in blue, which are obtained by following steps 1 and 2 for different initial conditions and yield a lower heat extraction.
		\end{enumerate*}
		}
	\label{fig:general}
\end{figure*}

\subsection{Setup} \label{subsec:general:setup}

A two-stroke refrigerator consists of three basic components:
	two reservoirs at different temperatures $T_\txtc$ and $T_\txth > T_\txtc$ and a controlled working system \cite{SeifertRepProgPhys2012,BenentiPhysRep2017}.
We start by developing our general scheme before moving on to specific applications in Secs.~\ref{sec:fridge} and \ref{sec:coh}.
The internal state of the working system is described by a vector of $N$ independent variables $\vec R_t$, which follows the time evolution equation
\begin{equation}\label{eq:gen:evolution}
	\dot{\vec R}_t = \vec F[\vec R_t, \omega_t]
\end{equation}
with dots indicating time-derivatives throughout.
The generator $\vec F$ thereby depends on the specific architecture of the device and it is assumed to be local in time, i.e., it only depends on the state vector $\vec R_t$ and the driving protocol $\omega_t$ at time $t$.
It may, however, be a non-linear function of these variables.
The external parameter $\omega_t$ plays a three-fold role;
	it controls the dynamics of the state vector, modulates the internal energy landscape of the working system, and it regulates the coupling to the reservoirs \cite{Gelbwaser-KlimovskyAdvancesInAtomicMolecularandOpticalPhysics2015}.

The key idea of our two-stroke scheme is to disentangle these effects.
To this end, we assume that the working system is connected either to the cold or the hot reservoir depending on whether $\omega_t$ is smaller or larger than a given threshold value $\switch\omega$.
A thermodynamic cooling cycle can then be realized as illustrated in Fig.~\ref{fig:introduction:cycle}.
In the work stroke, the control parameter $\omega_t$ changes continuously and does not exceed the threshold $\switch\omega$.
Thus, the working system is constantly coupled to the cold reservoir, from which it has picked up the heat
\begin{equation} \label{eq:gen:output}
	\mathcal Q_\txtc[\omega_t] \equiv \int_0^{\switch\period} Q[\vec R_t, \omega_t] \id t
\end{equation}
by the end of the stroke.
Here, $Q[\vec R_t, \omega_t]$ is the instantaneous heat flux flowing into the system.
Throughout this paper, we use calligraphic letters to denote functionals, which depend on the complete driving protocol $\omega_t$, for example the left hand side of \eqref{eq:gen:output}.
At the switching time $\switch\period$, $\omega_t$ is abruptly raised above the threshold $\switch\omega$.
This operation initializes the reset stroke, during which the control parameter follows a continuous trajectory without falling below $\switch\omega$.
Hence, the system is coupled to the hot reservoir throughout this stroke, which restores the initial state of the system and releases the heat
\begin{equation} \label{eq:gen:cost}
	\mathcal Q_\txth[\omega_t] \equiv -\int_{\switch\period}^{\period} Q[\vec R_t, \omega_t] \id t .
\end{equation}
The cycle is completed at the time $\period$ by instantaneously resetting the control parameter to its initial value.

The specific form of the function $Q[\vec R_t, \omega_t]$ is determined by the architecture of the refrigerator.
For example, if the working system can be described as an open quantum system in the weak coupling regime, this quantity can universally be identified as \cite{SpohnAdvChemPhys1978,AlickiJPhysA1979,GevaPhysRevE1994,VinjanampathyContempPhys2016}
\begin{equation} \label{eq:A}
	Q[\vec R_t, \omega_t] \equiv \tr[ H_t \dot\rho_t ] .
\end{equation}
Here, $H_t \equiv H[\omega_t]$ denotes the Hamiltonian of the working system and $\rho_t \equiv \rho[\vec R_t]$ the density matrix describing its state.
Remarkably, our two-stroke scheme enables a general optimization procedure even without such specifications, as we will show in the following.

\subsection{Maximum Heat Extraction} \label{subsec:general:output}

Our first aim is to find the control protocol $\outputX\omega_t$ that maximizes the heat extraction \eqref{eq:gen:output} for a given cycle time $\tau$.
To this end we proceed along the three steps illustrated in Fig.~\ref{fig:general}.

First, for the optimal work stroke, $\omega_t$ has to be chosen such that the extended objective functional for the extracted heat
\begin{equation} \label{eq:gen:ext_output}
	\mathcal Q_\txtc[\vec R_t, \vec \lambda_t, \omega_t] \equiv \int_0^{\switch\period} \!\bigl( Q[\vec R_t, \omega_t] - \vec \lambda_t \cdot ( \dot{\vec R}_t - \vec F[\vec R_t, \omega_t] ) \bigr) \id t
\end{equation}
becomes stationary, i.e., its functional derivative with respect to its arguments vanishes \cite{Kirk2004}.
Here, we have introduced a vector of Lagrange multipliers $\vec\lambda_t$ to account for the dynamical constraint \eqref{eq:gen:evolution}.
This extension of the parameter space makes it possible to treat the control parameter $\omega_t$ and the state $\vec R_t$ as independent variables.
Optimizing the functional \eqref{eq:gen:ext_output} is formally equivalent to applying the least-action principle in Hamiltonian mechanics with $\vec R_t$ and $\vec\lambda_t$ playing the role of generalized coordinates and canonical momenta, respectively \cite{Goldstein2002}.
The corresponding effective Hamiltonian is given by
\begin{equation} \label{eq:gen:H_work}
	\Xwork H[\vec R_t, \vec \lambda_t, \omega_t] \equiv Q[\vec R_t, \omega_t] + \vec \lambda_t\!\cdot\!\vec F[\vec R_t, \omega_t] .
\end{equation}
Thus, after fixing the initial conditions $\vec R_{t=0} = \vec R_0$ and $\vec\lambda_{t=0} = \vec\lambda_0$, the optimal protocol for the work stroke is uniquely determined by the canonical equations \cite{Kirk2004}
\begin{equation} \label{eq:gen:canonical_work}
	\dot{\vec R}_t = \frac{\partial \Xwork H}{\partial \vec \lambda_t} , \quad
	\dot{\vec \lambda}_t = -\frac{\partial \Xwork H}{\partial \vec R_t} \quad\text{and}\quad
	\frac{\partial \Xwork H}{\partial \omega_t} = 0 .
\end{equation}
Note that the last equation is purely algebraic.
Therefore, the initial value of the control parameter, $\omega_0$, is fixed by choosing $\vec R_0$ and $\vec\lambda_0$.

Second, since only the work stroke contributes to the extracted heat, the optimal reset stroke minimizes the reset time $\Xreset\durationFunctional \equiv \period - \switch\period$, during which the system returns to its initial state.
To implement this condition, we have to minimize the extended objective functional for the reset time
\begin{align}
	\Xreset\durationFunctional[\vec R_t, \vec\lambda_t, \omega_t]
		&\equiv \int_{\switch\period}^\period \!\bigl( 1 - \vec \lambda_t \cdot ( \dot{\vec R}_t - \vec F[\vec R_t, \omega_t] ) \bigr) \id t \nonumber \\
		&\equiv \int_{\switch\period}^\period \!\bigl( \Xreset H[\vec R_t, \vec\lambda_t, \omega_t] - \vec\lambda_t\!\cdot\!\dot{\vec R}_t \bigr) \id t \label{eq:gen:ext_duration}
\end{align}
with respect to the dynamical variables $\vec R_t$, $\vec\lambda_t$ and $\omega_t$, and the switching time $\switch\period$.
Thus, the optimal reset protocol can be found by solving the canonical equations
\begin{equation} \label{eq:gen:canonical_duration}
	\dot{\vec R}_t = \frac{\partial \Xreset H}{\partial \vec \lambda_t} , \quad
	\dot{\vec \lambda}_t = -\frac{\partial \Xreset H}{\partial \vec R_t} \quad\text{and}\quad
	\frac{\partial \Xreset H}{\partial \omega_t} = 0
\end{equation}
with respect to the boundary conditions
\begin{align} \label{eq:gen:bcs_duration}
	&\vec R_{t=\switch\period} = \switch{\vec R}[\vec R_0, \vec\lambda_0], \quad
	\vec R_{t=\period} = \vec R_0 \quad\text{and}\quad \\
	&\Xreset H[\vec R_{\switch\period}, \vec \lambda_{\switch\period}, \omega_{\switch\period}] = 0 . \nonumber
\end{align}
Here, $\switch{\vec R}$ is the state vector of the system after the optimal work stroke, and the end-point condition $\vec R_{t=\period} = \vec R_0$ replaces the initial condition for the Lagrange multipliers.

Note that the state $\vec R_t$ has to be continuous throughout the cycle \cite{EspositoEPL2010}, while the Lagrange multipliers $\vec\lambda_t$ of the work and reset strokes are independent variables; they therefore do not have to satisfy any boundary conditions.
The last requirement in \eqref{eq:gen:bcs_duration} minimizes $\Xreset\durationFunctional$ with respect to the initial time $\switch\period$ \cite{Kirk2004}.
In practice, the switching time $\switch\period$ and the initial Lagrange multipliers $\vec \lambda_{\switch\period}$ have to be determined together such that the conditions \eqref{eq:gen:bcs_duration} are satisfied.

The procedure above leads to the optimal protocol if the algebraic condition $\partial_{\omega_t} \Xreset H = 0$ can be satisfied throughout the reset stroke.
However, the reset Hamiltonian $\Xreset H$ does often not have a local extremum within the admissible range $[\switch\omega, \omega_\txtmax]$ of the control parameter \cite{DeffnerJPhysB2014,CavinaPhysRevA2018}.
The optimal reset protocol $\outputX\omega_t$ then has to assume one of the boundary values $\switch\omega$ or $\omega_\txtmax$, so that it minimizes the effective Hamiltonian $\Xreset H$.
Formally, we thus replace the last equation in \eqref{eq:gen:canonical_duration} by the more general requirement
\begin{equation} \label{eq:gen:pmp}
	\Xreset H[\outputX{\vec R}_t, \outputX{\vec\lambda}_t, \outputX\omega_t] \leq \Xreset H[\outputX{\vec R}_t, \outputX{\vec\lambda}_t, \omega] ,
\end{equation}
which is also known as Pontryagin's minimum principle \cite{Pontryagin1962,Kirk2004}.
Here, $\outputX{\vec R}_t$ and $\outputX{\vec\lambda}_t$ are the optimal trajectories of the state vector and the Lagrange multiplier, respectively.
The canonical equations \eqref{eq:gen:canonical_duration} can thus be integrated as follows.
First, for given initial conditions $\vec R_0$ and $\vec\lambda_0$, the initial value of the control parameter, $\omega_0$, has to be determined such that $\Xreset H[\vec R_0, \vec\lambda_0, \omega_0]$ becomes minimal.
If this function does not have a local minimum within the range $[\switch\omega, \omega_\txtmax]$, we either have $\omega_0 = \switch\omega$ or $\omega_0 = \omega_\txtmax$.
After fixing $\omega_0$, the state vector and the Lagrange multipliers can be propagated for a short time $\dt$ using the canonical equations.
The control parameter is then updated by minimizing the Hamiltonian $\Xreset H[\vec R_{\dt}, \vec\lambda_{\dt}, \omega_{\dt}]$ with respect to $\omega_{\dt}$.
Iterating this procedure until the final time $\tau$ yields the optimal trajectories $\outputX{\vec R}_t$, $\outputX{\vec \lambda}_t$ and $\outputX\omega_t$.
This prescription typically leads to protocols that are either constant or consist of constant pieces connected by continuous trajectories \cite{BoldtEPL2012}.
In Sec.~\ref{sec:fridge}, we will show how both of these cases can be handled in practice.

Third and finally, after completing steps 1 and 2, we arrive at the optimal protocol $\outputX\omega_t = \outputX\omega_t[\vec R_0, \vec\lambda_0]$ for fixed initial conditions $\vec R_0$ and $\vec\lambda_0$.
Inserting this solution into \eqref{eq:gen:output} renders the extracted heat an ordinary function of $2N$ variables, $\mathcal Q_\txtc = \mathcal Q_\txtc[\vec R_0, \vec\lambda_0]$.
The last step of our scheme thus consists of maximizing this function over the state space of the working system and the set of admissible Lagrange multipliers, i.e., those $\vec\lambda_0$, for which $\omega_0 = \omega_0[\vec R_0, \vec\lambda_0]$ falls into the permitted range $[\omega_\txtmin, \switch\omega]$.
We note that maximizing $\mathcal Q_\txtc$ over all initial conditions $\vec R_0$ and $\vec\lambda_0$ is equivalent to maximizing $\mathcal Q_\txtc$ over all switching times $\switch\period$ and all boundary values $\vec R_0$ and $\switch{\vec R}$, since these quantities are connected by a one-to-one mapping.

\subsection{Maximum Efficiency} \label{subsec:general:efficiency}

So far, we have developed a scheme to maximize the extracted heat per operation cycle of a general two-stroke refrigerator.
A thorough optimization of a thermal machine, however, also has to take into account the consumed input, which, for a cooling device, corresponds to the work $\mathcal W[\omega_t]$ that the external controller has to supply to drive the heat flux.
To this end, we now show how to find the optimal protocol $\effX\omega_t$, which maximizes the efficiency
\begin{align} \label{eq:fridge:efficiency}
	\eta[\omega_t] &\equiv \mathcal Q_{\txtc}[\omega_t] / \mathcal W[\omega_t] \\
		&= \mathcal Q_{\txtc}[\omega_t] / (\mathcal Q_{\txth}[\omega_t] - \mathcal Q_{\txtc}[\omega_t]) , \nonumber
\end{align}
a second key indicator for thermodynamic performance \cite{SeifertRepProgPhys2012}.
Note that here we have used the first law of thermodynamics to express the work input $\mathcal W[\omega_t]$ in terms of the released and the extracted heat, $\mathcal Q_\txth[\omega_t]$ and $\mathcal Q_\txtc[\omega_t]$.
Owing to the second law, the figure of merit \eqref{eq:fridge:efficiency} is subject to the Carnot bound
\begin{equation} \label{eq:carnot}
	\eta[\omega_t] \leq \eta_C \equiv \frac{T_\txtc}{T_\txth - T_\txtc} ,
\end{equation}
which is saturated in the reversible limit at the price of vanishing cooling power \cite{NiskanenPhysRevB2007}.
Hence, for a practical optimization criterion, we have to fix both the cycle time $\period$ and the heat extraction $\mathcal Q_\txtc[\omega_t] = \fixedOutput$.
Maximizing the efficiency \eqref{eq:fridge:efficiency} then amounts to minimizing the effective input $\mathcal Q_\txth[\omega_t]$, i.e., the average heat injected into the hot reservoir per operation cycle.

The corresponding protocol $\effX\omega_t = \effX\omega_t[\fixedOutput]$ renders the work stroke as short as possible such that the maximum amount of time is left to reduce the heat release in the reset stroke \footnote{
		An alternative way to see that $\effX\omega_t[\vec R_0, \vec\lambda_0] = \outputX\omega_t[\vec R_0, \vec\lambda_0]$ must hold during the work stroke is to observe that the protocol minimizing the effective input $\mathcal Q_\txth$ for fixed effective output $\fixedOutput$ simultaneously maximizes the output for fixed input.
	}.
Hence, in the first step, we have to minimize the working time
\begin{align} \label{eq:gen:ext_duration_eff}
	&\Xwork\durationFunctional[\vec R_t, \vec\lambda_t, \omega_t, \mu] \\
	&\equiv \int_0^{\switch\period} \!\bigl( 1 - \mu(\fixedOutput + \vec\lambda_t\!\cdot\!\dot{\vec R}_t - \Xwork H[\vec R_t, \vec\lambda_t, \omega_t] ) \bigr) \id t , \nonumber
\end{align}
where the time-independent Lagrange multiplier $\mu$ has been introduced to fix the total heat extraction $\fixedOutput$.
This variational problem again leads to the canonical equations \eqref{eq:gen:canonical_work}, which have to be solved for given initial conditions $\vec R_0$ and $\vec\lambda_0$ to find the optimal work protocol.
In fact, this protocol also maximizes the heat extraction for every given time $t$, i.e., we have $\effX\omega_t[\vec R_0, \vec\lambda_0] = \outputX\omega_t[\vec R_0, \vec\lambda_0]$ during the work stroke.
However, the switching time $\switch\period = \switch\period[\vec R_0, \vec\lambda_0, \fixedOutput]$ now has to be chosen such that the constraint
\begin{equation} \label{eq:gen:switching_time_eff}
	\int_0^{\switch\period} Q[\vec R_t, \omega_t] \id t = \fixedOutput
\end{equation}
is satisfied.
Hence, the switching time is now determined by the work stroke rather than the reset stroke.

After completing step 1, the optimal reset protocol is found by minimizing the functional
\begin{equation} \label{eq:gen:ext_cost}
	\mathcal Q_\txth[\vec R_t, \vec\lambda_t, \omega_t]
		\equiv \int_{\switch\period}^\period \!\bigl( -Q[\vec R_t, \omega_t] - \vec\lambda_t \cdot ( \dot{\vec R}_t - \vec F[\vec R_t, \omega_t] ) \bigr) \id t
\end{equation}
for the boundary conditions
\begin{equation} \label{eq:gen:bcs_cost}
	\vec R_{t=\switch\period} = \switch{\vec R}[\vec R_0, \vec\lambda_0, \fixedOutput] \quad\text{and}\quad
	\vec R_{t=\period} = \vec R_0 .
\end{equation}
This problem will, depending on the initial conditions, only admit a proper solution if the device can actually produce the cooling power $\fixedOutput / \period$ in a cyclic mode of operation.
It might therefore be helpful to introduce an intermediate step, which decides whether or not the cycle can be closed for the boundary conditions \eqref{eq:gen:bcs_cost}.
To solve the canonical equations for the objective functional \eqref{eq:gen:ext_cost}, it might again be necessary to invoke Pontryagin's minimum principle, as we will demonstrate explicitly in Sec.~\ref{subsec:sc:efficiency}.

Once the reset protocol has been determined, the efficiency \eqref{eq:fridge:efficiency} can be reduced to an ordinary function of $\vec R_0$ and $\vec\lambda_0$.
Maximizing this function under the constraint $\omega_0[\vec R_0, \vec\lambda_0] \in [\omega_\txtmin, \switch\omega]$ yields the maximal-efficiency protocol $\effX\omega_t[\fixedOutput]$.
Note that the set of admissible initial conditions is thereby also restricted by fixing the heat extraction $\fixedOutput$.

\section{Quantum Microcooler I -- Semiclassical Regime} \label{sec:fridge}

\subsection{System} \label{subsec:fridge:system}

\begin{figure}
	\centering
	\sidesubfloat[]{
		\includegraphics[width=0.85\linewidth]{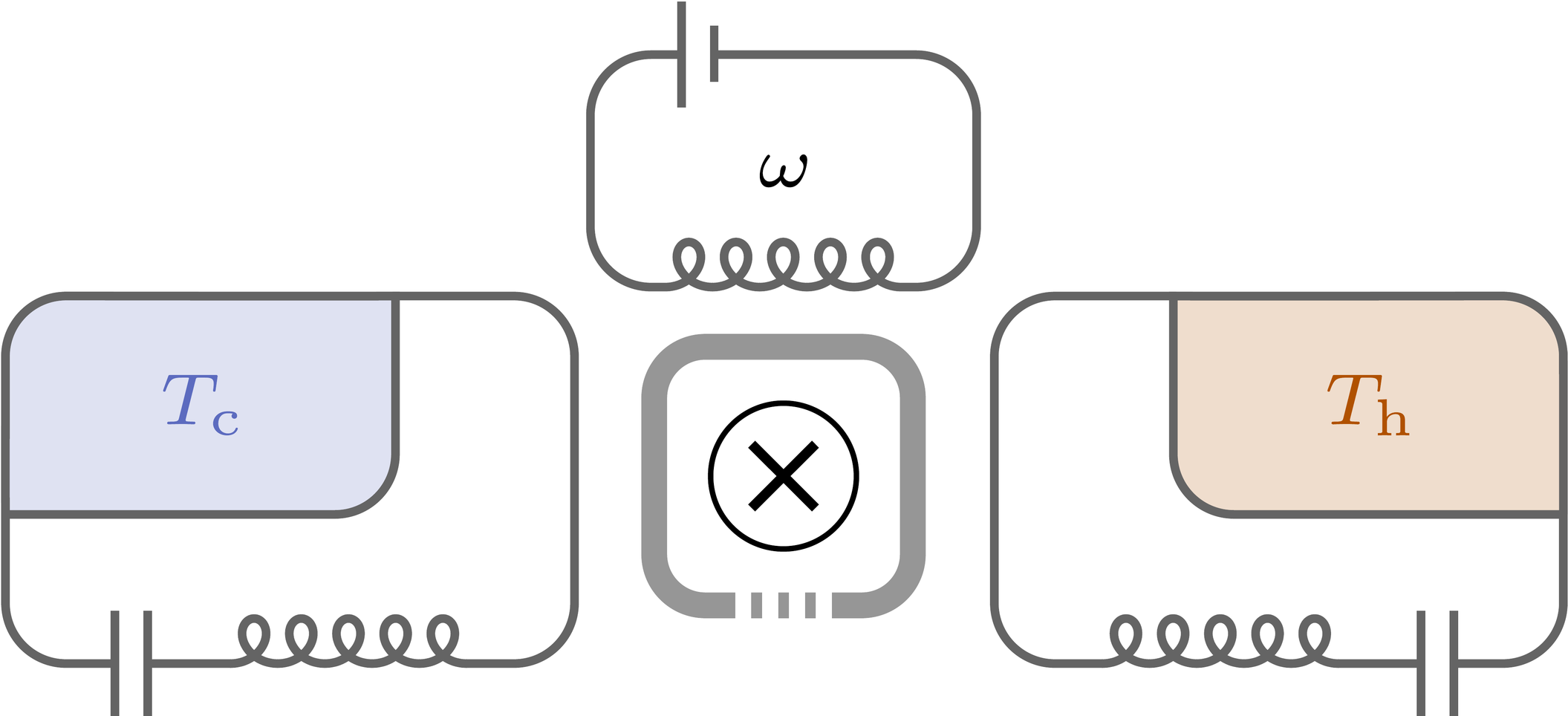}
		\label{fig:fridge:setup}
	}\\[2em]
	\sidesubfloat[]{
		\includegraphics[width=0.85\linewidth]{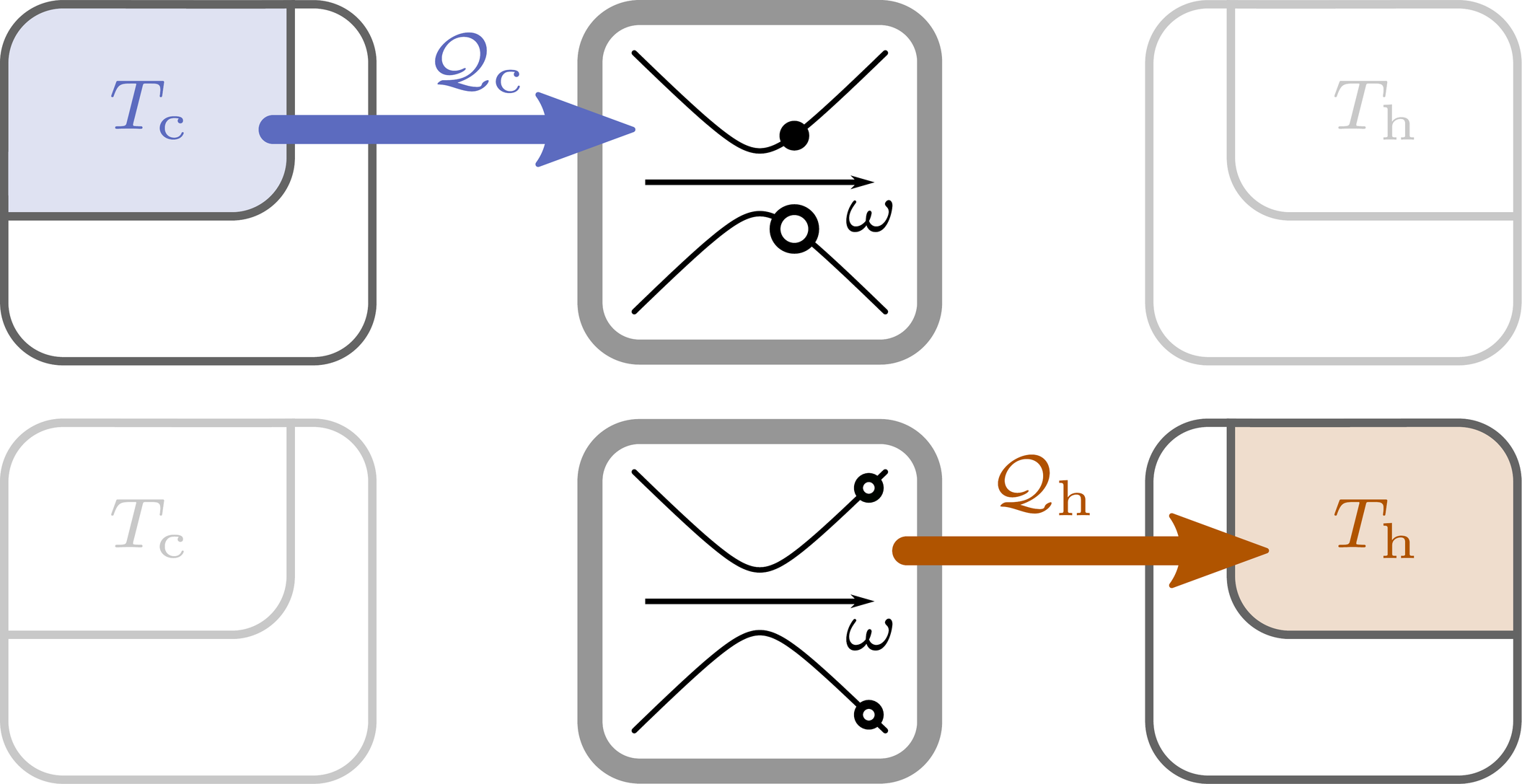}
		\label{fig:fridge:cycle}
	}
	\caption[this_is_required]{Quantum microcooler.
		\begin{enumerate*}[(a)]
			\item Sketch of the experimental setup described in Refs.~\cite{NiskanenPhysRevB2007,KarimiPhysRevB2016,RonzaniNatPhys2018}.
				A superconducting qubit is coupled to two resonant circuits with different resonance frequencies.
				Each circuit contains a metallic island acting as a mesoscopic reservoir with temperature $T_\txtc$ and $T_\txth > T_\txtc$, respectively.
				An additional bias circuit is used to control the level splitting of the qubit by varying the applied magnetic flux.
			\item Scheme of the thermodynamic cooling cycle.
				The two central diagrams show the energy levels of the qubit as a function of the external bias $\omega$ and the corresponding populations at the beginning of each stroke.
				By the end of the work stroke, the qubit has picked up the heat $\mathcal Q_\txtc$ from the cold island.
				The level splitting is then instantaneously increased to tune the qubit into resonance with the hot island.
				During the following reset stroke, the initial level populations are restored, while the heat $\mathcal Q_\txth$ flows into the hot reservoir.
				The cycle is completed by setting the level splitting back to its initial value, thus reconnecting the qubit to the cold island.
		\end{enumerate*}}
	\label{fig:fridge}
\end{figure}

We will now show how our general theory can be applied to a concrete problem of quantum engineering.
Specifically, we optimize the performance of a quantum microcooler, which can be implemented with superconducting components, see Fig.~\ref{fig:fridge:setup}.
The core of this device is an engineered two-level system with Hamiltonian \cite{NiskanenPhysRevB2007}
\begin{equation} \label{eq:fridge:hamiltonian}
	H_t \equiv \frac{\hbar\Delta}{2} \sigma_x + \frac{\hbar\omega_t}{2} \sigma_z .
\end{equation}
Here, $\hbar$ denotes the reduced Planck constant, $\sigma_x$ and $\sigma_z$ are Pauli matrices, $\Delta$ corresponds to the device-specific tunneling energy and $\omega_t$ is the tunable energy bias, which plays the role of the external control parameter.
This system is embedded in an electronic circuit, which couples it either to a cold or a hot reservoir depending on the value of $\omega_t$.
Thus, applying a suitable periodic control protocol $\omega_t$ makes it possible to realize a two-stroke cooling cycle, as illustrated in Fig.~\ref{fig:fridge:cycle}.

\subsection{Step-Rate Model} \label{subsec:sc:model}

For a quantitative description of the microcooler, we consider the model shown in Fig.~\ref{fig:fridge:setup}, which makes it possible to determine the optimal control protocol analytically.
To this end, we here focus on the semiclassical limit, where the tunneling energy $\Delta$ is negligible and the Hamiltonian commutes with itself at different times.
The periodic density matrix of the working system is then fully determined by the level populations and can be parametrized as
\begin{equation} \label{eq:sc:param}
	\rho_t \equiv \frac 1 2 ( \mathbbm 1 + R_t\, \sigma_z ) .
\end{equation}
The state variable $R_t$ thereby obeys the Bloch equation \cite{GevaPhysRevE1994}
\begin{align} \label{eq:sc:bloch}
	\dot R_t = F[R_t, \omega_t] &\equiv -\Gamma^+[\omega_t]\, R_t - \Gamma^-[\omega_t] \quad\text{with} \\
		\Gamma^\pm[\omega_t] &\equiv \gamma[\omega_t] \bigl( 1 \pm \exp[-\hbar\omega_t / T[\omega_t]] \bigr) . \nonumber
\end{align}
Here, the Boltzmann factors appear due to the detailed balance condition, which fixes the relative frequency of thermal excitation and relaxation events \cite{SeifertRepProgPhys2012}.
The corresponding temperature is determined by the reservoir coupled to the system, i.e.,
\begin{equation} \label{eq:fridge:temp}
	T[\omega \leq \switch\omega] \equiv T_\txtc \quad\text{and}\quad T[\omega > \switch\omega] \equiv T_\txth ,
\end{equation}
where $\switch\omega$ corresponds to the threshold energy of the device.
Note that Boltzmann's constant is set to $1$ throughout.
The factor $\gamma[\omega]$ in \eqref{eq:sc:bloch} accounts for the finite energy range of the coupling mechanism between working system and reservoirs, which depends on the specific design of the circuit.
For the sake of simplicity, we here use an idealized model, where the rates \eqref{eq:sc:bloch} feature a step-type dependence on $\omega$, i.e., we set
\begin{equation} \label{eq:fridge:gamma}
	\gamma[\omega] \equiv \gamma = \text{const.\ for } 0 < \omega \leq \omega_\txtmax
\end{equation}
and $\gamma[\omega] \equiv 0$ otherwise.
Hence, the two-level system is decoupled from its environment if $\omega$ falls outside its admissible range.
Note that we have set $\omega_\txtmin$ to zero.

Under weak-coupling conditions, the instantaneous heat flux into the qubit is given by \eqref{eq:A}.
The average amount of heat that the microcooler extracts from the cold reservoir in one cycle of duration $\tau$ then becomes
\begin{equation} \label{eq:sc:cooling_power}
	\mathcal Q_\txtc[\omega_t]
		\equiv \int_0^{\switch\period} \frac{\hbar\omega_t}{2} \dot R_t \id t
		= \int_0^{\switch\period} \frac{\hbar\omega_t}{2} F[R_t, \omega_t] \id t ,
\end{equation}
where $\switch\period \in [0, \period]$ denotes the length of the work stroke.
Accordingly, the average heat injected into the hot reservoir is given by
\begin{equation}
	\mathcal Q_\txth[\omega_t] \equiv -\int_{\switch\period}^\period \frac{\hbar\omega_t}{2} F[R_t, \omega_t] \id t .
\end{equation}

\subsection{Maximum Heat Extraction} \label{subsec:sc:cooling_power}

\captionsetup[subfigure]{singlelinecheck=off,justification=raggedright,position=top}
\begin{figure*}%
	\centering%
	\null\hfill%
	\subfloat[]{%
		\centering%
		\includegraphics[scale=1]{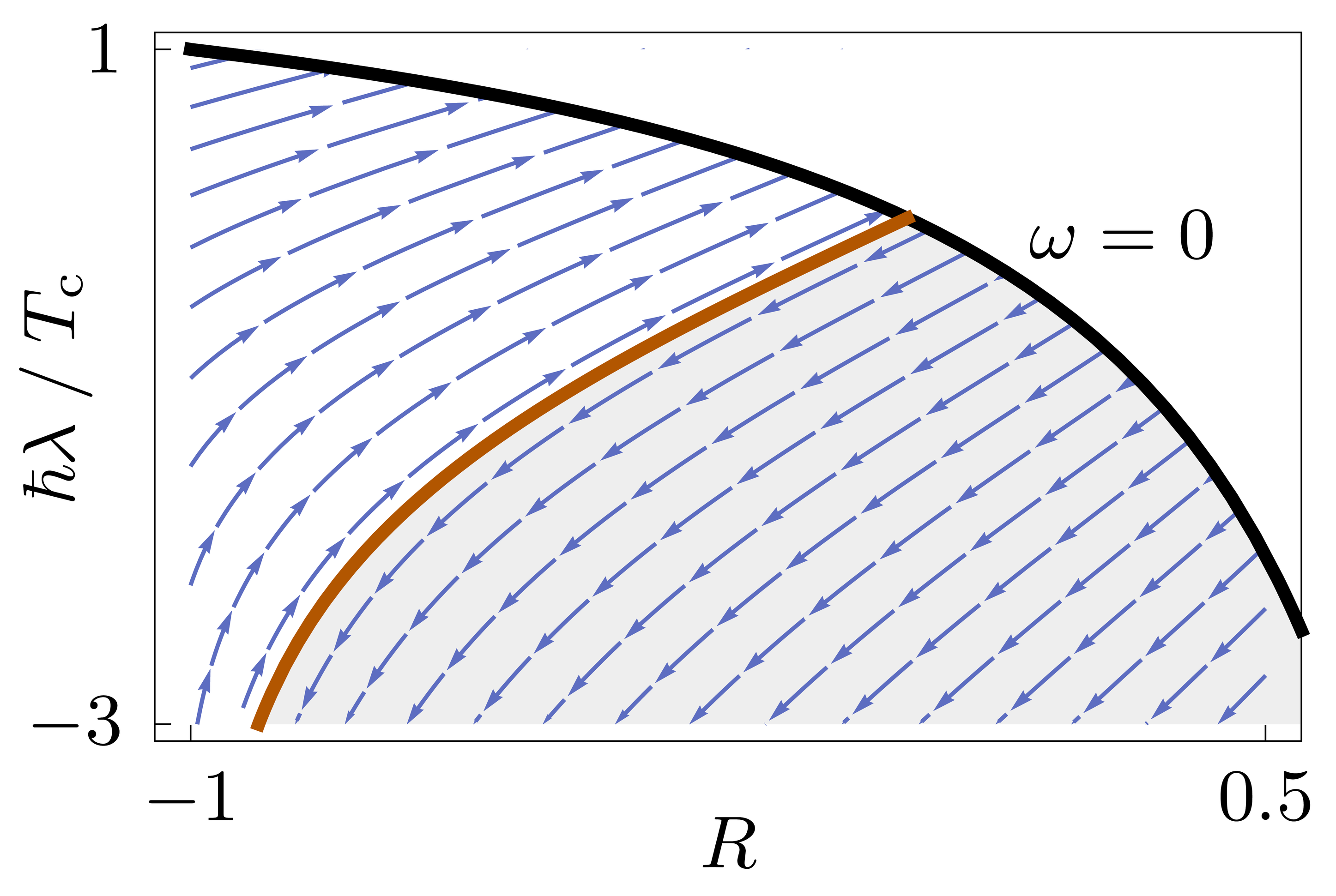}%
		\label{fig:sc_optimization:step1}%
	}\hfill%
	\subfloat[]{%
		\centering%
		\includegraphics[scale=1]{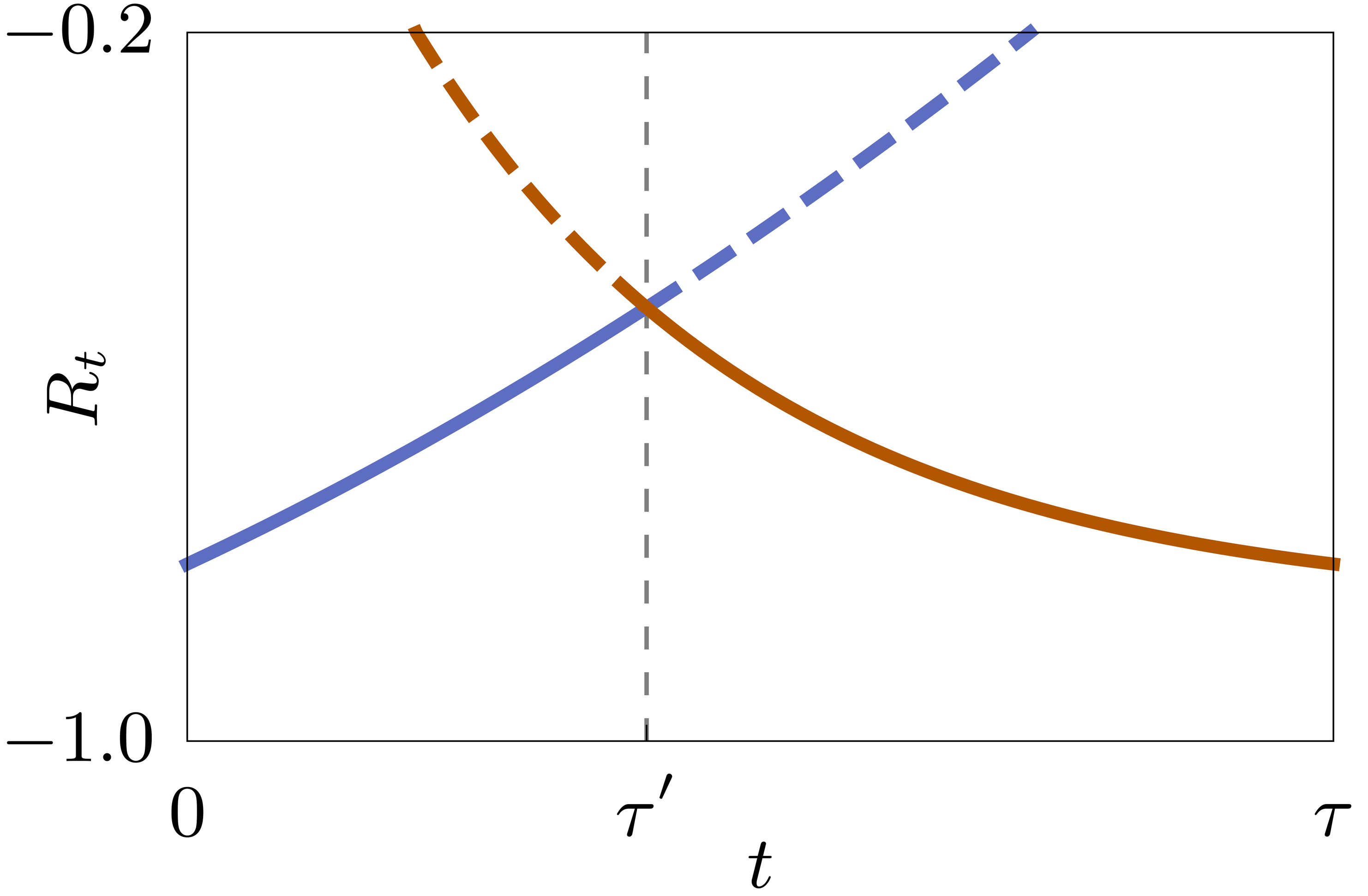}%
		\label{fig:sc_optimization:step2}%
	}\hfill%
	\subfloat[]{
		\centering%
		\includegraphics[scale=1]{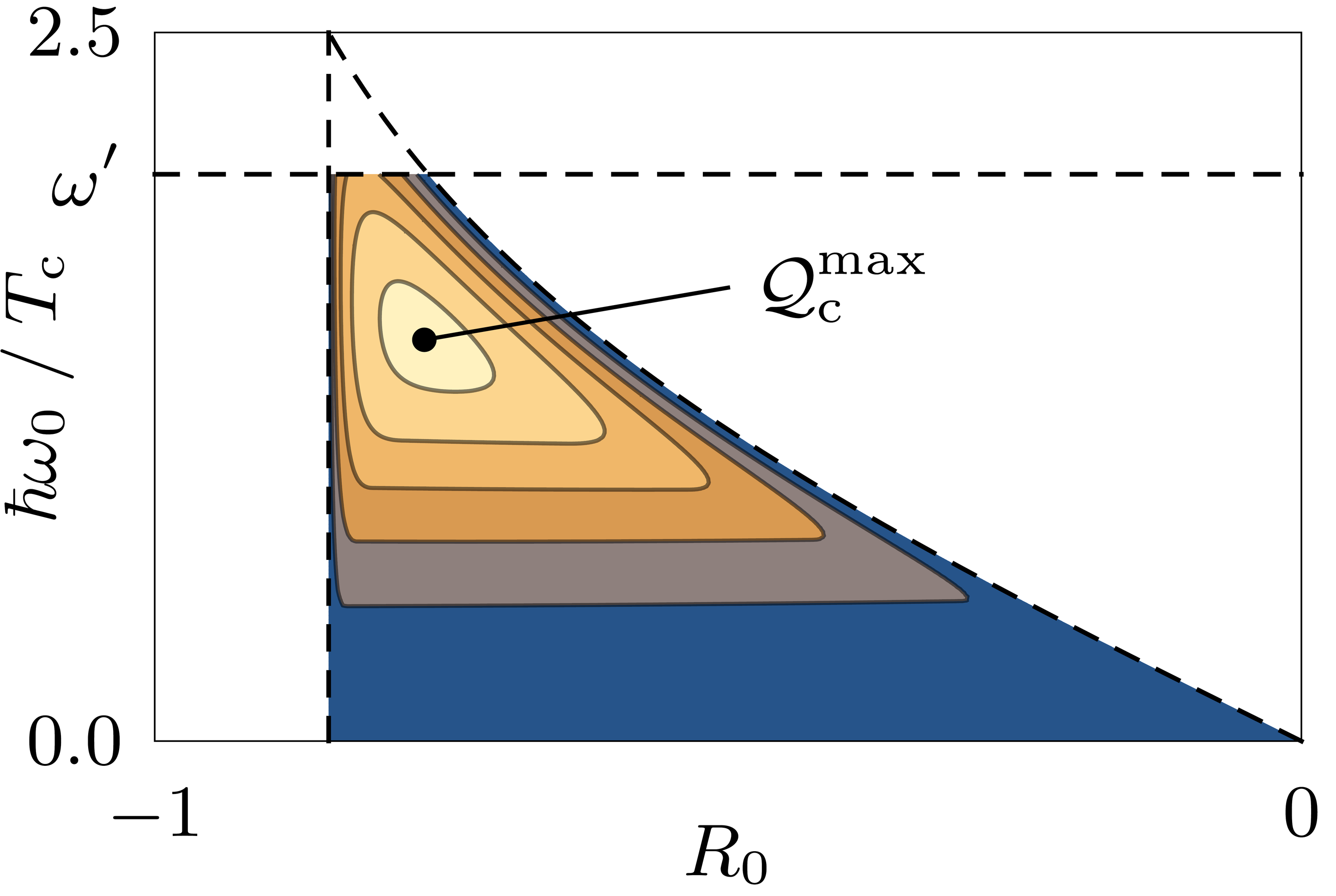}%
		\label{fig:sc_optimization:step3}%
	}%
	\hfill\null%
	\caption[this_is_required]{Maximizing the cooling power of a quantum microcooler in three steps.
		\begin{enumerate*}[(a)]
		\item The plot shows the flow of the effective Hamiltonian vector field \eqref{eq:sc:canonical_Phi}, which determines the optimal dynamics of the system during the work stroke.
			The black line marks the boundary of the physical region of the effective phase space, where the level splitting $\omega$ is positive.
			The red line separates solutions with positive (unshaded) and negative (shaded) cooling power.
		\item The state trajectories during the optimal work and reset stroke, \eqref{eq:sc:solution_R} and \eqref{eq:sc:hot_stroke}, are plotted in blue and red, respectively, for typical initial conditions.
			Their intersection point determines the optimal switching time $\switch\period$.
			The solid line shows the combined optimal state trajectory, which excludes the dashed parts.
		\item The maximum heat extraction $\mathcal Q_\txtc[R_0, \omega_0]$ is plotted over the admissible range \eqref{eq:sc:ic_restrictions_final} of initial conditions, which is bounded by the dashed black curves corresponding to $R_0 = -\tanh[\hbar\omega_0 / (2T_\txtc)]$, $R_0 = -\tanh[\hbar\omega_\txtmax / (2T_\txth)]$ and $\omega_0 = \switch\omega$.
			Brighter colors indicate a larger amount of extracted heat.
			The global maximum $\mathcal Q_\txtc^\txtmax$ is shown with a dot.
		\end{enumerate*}
		All panels were created with the parameter values $\switch\omega = 2\, T_\txtc / \hbar$, $\omega_\txtmax = 5\, T_\txtc / \hbar$, $T_\txth = 2\, T_\txtc$ and $\period = 3 / \gamma$.
		}%
\end{figure*}

The extracted heat \eqref{eq:sc:cooling_power} can be maximized using the general scheme of Sec.~\ref{subsec:general:output}.
To this end, we first have to determine the optimal work stroke, which is described by the effective Hamiltonian \footnote{
	Note that we have rescaled the objective functional, giving the Lagrange multiplier the same dimension as $\omega$.}
\begin{equation} \label{eq:sc:output_Hw}
	\Xwork H[R_t, \lambda_t, \omega_t] = -(\omega_t + \lambda_t) (\Gamma^+[\omega_t] R_t + \Gamma^-[\omega_t]) .
\end{equation}
The corresponding canonical equations follow from \eqref{eq:gen:canonical_work} and are given by
\begin{align} \label{eq:sc:canonical}
	\dot R_t &= -\Gamma^+[\omega_t]\, R_t - \Gamma^-[\omega_t] , \\
	\dot\lambda_t &= \Gamma^+[\omega_t] \left( \omega_t + \lambda_t \right) \quad\text{and} \nonumber \\
	\omega_t &= (T_\txtc / \hbar) - \lambda_t - (T_\txtc / \hbar)\, W_0\!\left[ \e^{1 - \hbar\lambda_t / T_\txtc} \frac{1+R_t}{1-R_t} \right] , \nonumber
\end{align}
where we have explicitly solved the last equation for $\omega_t$.
We used that $\gamma[\omega_t] = \gamma$ and $T[\omega_t] = T_\txtc$ throughout the work stroke and $W_0$ denotes the upper branch of the Lambert $W$ function, which is defined as the solution to
\begin{equation} \label{eq:lambert}
	x \equiv W[x] \e^{W[x]} \quad\text{for}\quad x \geq -1 / \e .
\end{equation}
Upon eliminating $\omega_t$, the canonical equations \eqref{eq:sc:canonical} reduce to an autonomous system of first-order differential equations,
\begin{equation} \label{eq:sc:canonical_Phi}
	\vectwo{\dot R_t}{\dot\lambda_t} = \vectwo{\Phi_R[R_t, \lambda_t]}{\Phi_\lambda[R_t, \lambda_t]} \equiv \vec\Phi[R_t, \lambda_t]
\end{equation}
The flow of the Hamiltonian vector field $\vec\Phi[R, \lambda]$ is plotted in Fig.~\ref{fig:sc_optimization:step1}.
As a key observation, we find that the sign of $\dot R_t = \Phi_R[R_t, \lambda_t]$, which determines the direction of the instantaneous heat flux $Q_t = \hbar\omega_t \dot R_t / 2$, does not change along the optimal trajectories.
Hence, since our aim is to maximize the heat extraction from the cold reservoir, the initial values $R_0$ and $\lambda_0$ have to be chosen such that
\begin{equation} \label{eq:sc:ic_restriction}
	\dot R_0 = \Phi_R[R_0, \lambda_0] > 0 .
\end{equation}
Solving \eqref{eq:sc:canonical_Phi} under this condition and inserting the result into the third canonical equation \eqref{eq:sc:canonical} yields the protocol
\begin{equation} \label{eq:sc:solution_w}
	\outputX\omega_t = \frac{T_\txtc}{\hbar} \log\left[ \frac{2 - 2 C_1\, W_{-1}[C_2 e^{-\gamma t}]}{C_1\, W_{-1}[C_2 e^{-\gamma t}]^2} - 1 \right]
\end{equation}
and the corresponding state trajectory
\begin{equation} \label{eq:sc:solution_R}
	\outputX R_t = C_1 \left( 1 + W_{-1}[C_2 e^{-\gamma t}] \right)^2 - C_1 - 1 .
\end{equation}
Here, $W_{-1}$ denotes the lower branch of the Lambert W function and the constants $C_1$ and $C_2$ can be expressed in terms of the initial values $R_0$ and $\omega_0$, see Appendix~\ref{sec:appendix1}.
We note that the results \eqref{eq:sc:solution_w} and \eqref{eq:sc:solution_R} can also be obtained using a brute-force approach, where the dynamical constraint \eqref{eq:sc:bloch} is solved explicitly rather than being enforced through a Lagrange multiplier. (For further details see Appendix~\ref{sec:appendix1}.)
However, this approach crucially relies on the one-to-one correspondence \eqref{eq:sc:bloch} between the derivative $\dot R_t$ of the state variable and the control parameter $\omega_t$.
It is therefore not generally applicable.

To close the optimal cycle, the reset stroke has to restore the initial state $R_0$ of the system in minimal time.
According to Pontryagin's principle, the corresponding protocol can be found by minimizing the effective Hamiltonian
\begin{equation} \label{eq:sc:hamiltonian_reset}
	\Xreset H[R_t, \lambda_t, \omega_t] = 1 + \lambda_t\, F[R_t, \omega_t]
\end{equation}
with respect to $\omega_t$.
The variables $R_t$ and $\lambda_t$ thereby have to obey the canonical equations
\begin{equation} \label{eq:sc:canonical_reset}
	\dot R_t = F[R_t, \omega_t] \quad\text{and}\quad \dot\lambda_t = \Gamma^+[\omega_t]\, \lambda_t
\end{equation}
and the additional constraint
\begin{equation} \label{eq:sc:ic_reset}
	\Xreset H[R_{\switch\period}, \lambda_{\switch\period}, \omega_{\switch\period}] = 0
\end{equation}
at the yet undetermined optimal switching time $\switch\period$.

This problem can be approached as follows.
First, we observe that \eqref{eq:sc:ic_reset} implies
\begin{equation}
	\lambda_{\switch\period} = -1 / F[R_{\switch\period}, \omega_{\switch\period}] = -1 / \dot R_{\switch\period} .
\end{equation}
Since $R_t$ increases monotonically during the work stroke, it has to decrease during the reset.
Consequently, we have to choose $\lambda_{\switch\period} > 0$.
Minimizing the effective Hamiltonian \eqref{eq:sc:hamiltonian_reset} at the switching time $\switch\period$ is then equivalent to minimizing $F[R_{\switch\period}, \omega_{\switch\period}]$.
Second, the generator $F$ is a monotonically decreasing function of $\omega_{\switch\period}$ for any admissible value of $R_{\switch\period}$.
Thus, it follows that $\omega_{\switch\period} = \omega_\txtmax$, i.e., the control parameter abruptly jumps to its maximum at the beginning of the reset stroke.
Third, owing to \eqref{eq:sc:canonical_reset}, the sign of the Lagrange multiplier is conserved along its optimal trajectory.
Therefore, the same argument applies at any later time $t > \switch\period$ and we can conclude that $\outputX\omega_t = \omega_\txtmax$ throughout the reset stroke.
We note that this result could have been inferred directly from the Bloch equation \eqref{eq:sc:bloch} and the observation $\partial_{\omega_t} F[R_t, \omega_t] < 0$, which entails that the reset can always be accelerated by increasing $\omega_t$.
However, here we have chosen to follow the formal scheme of Sec.~\ref{sec:general} to illustrate the use of Pontryagin's principle.

Finally, we have to make sure that the state $R_t$ is continuous throughout the cycle.
To this end, its trajectory during the reset stroke,
\begin{equation} \label{eq:sc:hot_stroke}
	\outputX R_t = R_0\, \e^{\Gamma^+ (\period - t)} + (\Gamma^- / \Gamma^+) \left( \e^{\Gamma^+ (\period - t)} - 1 \right)
\end{equation}
with $\Gamma^\pm \equiv \Gamma^\pm[\omega_\txtmax]$, has to match the optimal work-stroke trajectory \eqref{eq:sc:solution_R} at $\switch\period$, see Fig.~\ref{fig:sc_optimization:step2}.
Numerically solving this condition yields the switching time $\switch\period$ and completes the optimal protocol $\outputX\omega_t[R_0, \omega_0]$ \footnote{
		Since the work stroke has a maximum duration, there are admissible initial conditions for which there is no intersection.
		We handle this case in practice by allowing for an intermediate time in which the system is decoupled from both reservoirs and the system state remains constant.
		We find that the protocol yielding the maximal average cooling power never decouples the system from both reservoirs.
	}.
Inserting this protocol back into the functional \eqref{eq:sc:cooling_power} together with \eqref{eq:sc:solution_R} and \eqref{eq:sc:hot_stroke} gives the maximal heat extraction $\mathcal Q_\txtc[R_0, \omega_0]$.

This function must now be maximized over the admissible range of initial values $R_0$ and $\omega_0$, which is restricted by the conditions
\begin{align} \label{eq:sc:ic_restrictions_final}
	R_0 &< -\tanh\left[ \hbar\omega_0 / (2T_\txtc) \right] , \\
	R_0 &> -\tanh\left[ \hbar\omega_\txtmax / (2T_\txth) \right] \nonumber
\end{align}
and the requirement that $\omega_0 \leq \switch\omega$, see Fig.~\ref{fig:sc_optimization:step3}.
The constraints \eqref{eq:sc:ic_restrictions_final} follow from \eqref{eq:sc:ic_restriction} and \eqref{eq:sc:hot_stroke}, respectively.
They ensure that the heat extraction $\mathcal Q_\txtc[R_0, \omega_0]$ is positive and that the initial state of the system can be restored during the reset.
To determine the maximal extracted heat $\mathcal Q_\txtc^\txtmax$ and the corresponding initial values, we employ a constrained optimization algorithm \cite{ZhuACMTransMathSoftw1997}, which finds $\mathcal Q_\txtc^\txtmax$ either inside the admissible range \eqref{eq:sc:ic_restrictions_final} or on the boundary $\omega_0 = \switch\omega$.

\begin{figure}
	\centering
	\sidesubfloat[]{
		\includegraphics[scale=1]{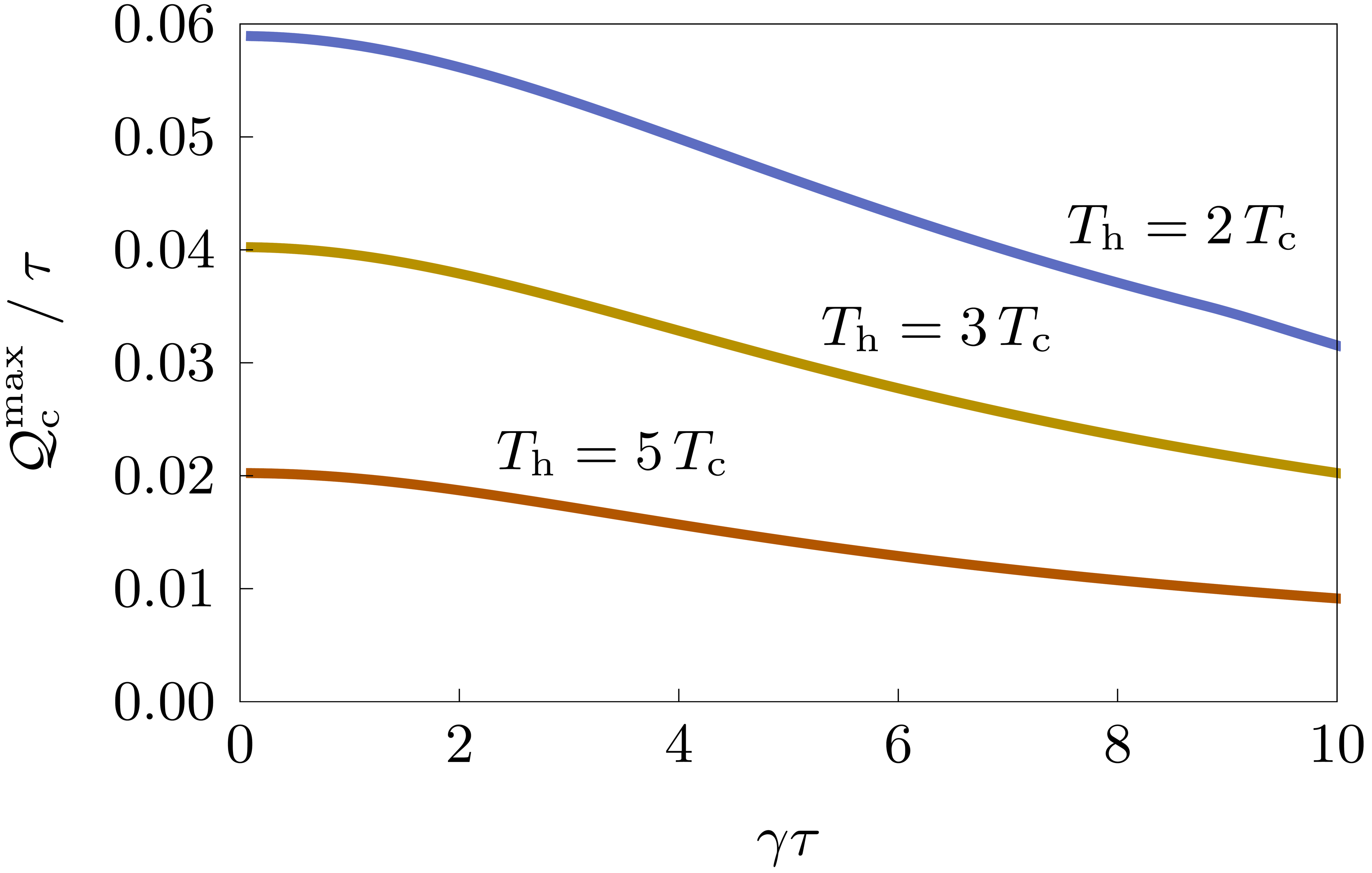}
		\label{fig:sc:results}
	}\\[1em]
	\sidesubfloat[]{
		\includegraphics[scale=1]{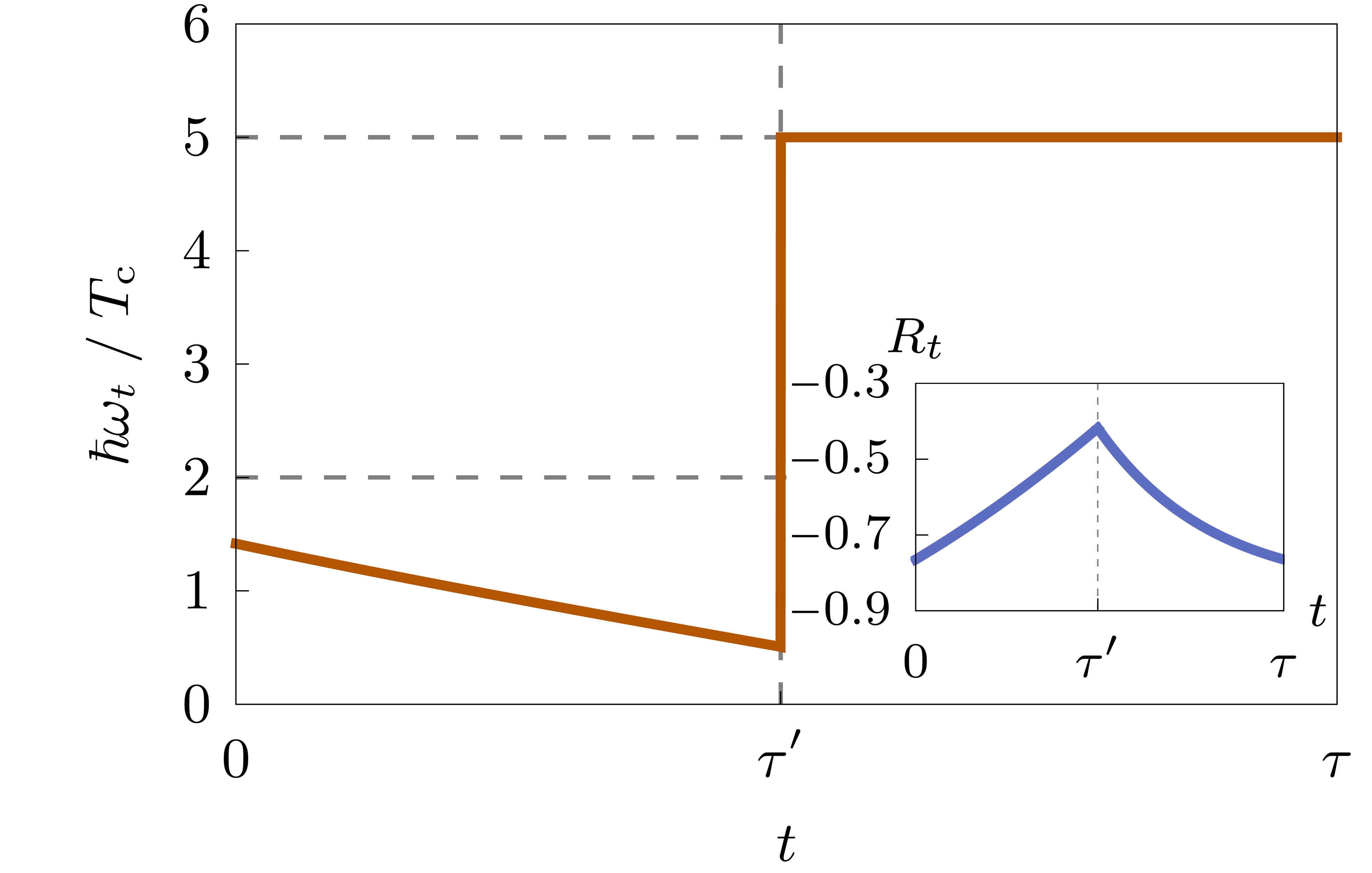}
		\label{fig:sc:protocol}
	}
	\caption[necessary]{Microcooler at optimal cooling power.
		\begin{enumerate*}[(a)]
		\item Maximum cooling power in units of $\gamma T_\txtc$ as a function of the dimensionless cycle time $\gamma\tau$ for different temperatures of the hot reservoir $T_\txth$.
		\item Optimal control protocol for $T_\txth = 2\, T_\txtc$ and $\period = 2\, \gamma^{-1}$.
			The inset shows the corresponding trajectory of the state variable $R_t$.
		\end{enumerate*}
		Here, we have used $\switch\omega = 2\, T_\txtc / \hbar$ and $\omega_\txtmax = 5\, T_\txtc / \hbar$.}
	\label{fig:sc:results_full}
\end{figure}

Figure~\ref{fig:sc:results} summarizes the results of this section.
The first plot shows the optimal cooling power $\mathcal Q_\txtc^\txtmax / \period$ as a function of the cycle time $\period$ for different values of the high temperature $T_\txth$.
We find that $\mathcal Q_\txtc^\txtmax / \period$ generally decreases with $\tau$.
Hence, for a large cooling power, the device must be operated fast.
For similar recent findings, the reader may consult Refs.~\cite{ErdmanArXiv181205089Quant-Ph2018,PekolaArXiv181210933Quant-Ph2018}.
Furthermore, the cooling power becomes successively smaller as $T_\txth$ increases.
This result confirms the natural expectation that the microcooler becomes less effective when it has to work against a larger temperature gradient.

Figure~\ref{fig:sc:protocol} illustrates the general behavior of our model during the optimal cycle.
In the work stroke, the state variable $\outputX R_t$ monotonically increases, while the control parameter $\outputX \omega_t$ monotonically decreases until the switching time is reached; at this point, no more heat can be extracted from the cold reservoir in a cyclic mode of operation, i.e., the work stroke has reached the maximal length.
In the reset stroke, the control parameter is constantly at its maximum, while $R_t$ returns to its initial value following an exponential decay.

\subsection{Maximum Efficiency} \label{subsec:sc:efficiency}

Having maximized the extracted heat of our microcooler mod\-el, we now focus on its thermodynamic efficiency \eqref{eq:fridge:efficiency}.
The optimal protocol $\effX\omega_t[\fixedOutput]$, which maximizes this figure of merit for a fixed heat extraction $\fixedOutput$, can be found using the scheme developed in Sec.~\ref{subsec:general:efficiency}.
During the work stroke, we have $\effX\omega_t[R_0, \omega_0] = \outputX\omega_t[R_0, \omega_0]$, that is, for $0 \leq t \leq \switch\period$ and fixed initial values $R_0$ and $\omega_0$, the protocol $\effX\omega_t[\fixedOutput]$ is given by \eqref{eq:sc:solution_w}.
The switching time $\switch\period$ can thus be determined from the constraint
\begin{equation}
	\int_0^{\switch\period} \frac{\hbar\omega_t}{2} F[R_t, \omega_t] \id t = \mathcal Q_\txtc^\ast
\end{equation}
using \eqref{eq:sc:solution_w} and \eqref{eq:sc:solution_R}.

The optimal reset stroke has to restore the initial state of the system while at the same time minimizing the dissipated heat $\mathcal Q_\txth[\omega_t]$.
To this end, the control protocol has to be chosen such that the effective Hamiltonian
\begin{equation} \label{eq:sc:efficiency_Hr}
	\Xreset H[R_t, \lambda_t, \omega_t] = (\omega_t + \lambda_t) (\Gamma^+[\omega_t] R_t + \Gamma^-[\omega_t])
\end{equation}
becomes minimal at every time $\switch\period \leq t \leq \period$, while $\lambda_t$ and $R_t$ obey the corresponding canonical equations.
For a given initial value $\lambda_0$ of the Lagrange multiplier, this problem can be solved using the procedure described in Sec.~\ref{subsec:general:efficiency}.
However, the situation is in practice complicated by the fact that $\lambda_0$ is determined only implicitly by the end-point condition $R_\period = R_0$.
It would still be possible to carry out the iteration scheme for every admissible value $\lambda_0$ and then pick the optimal protocol that closes the cycle.
This approach can, however, be expected to be numerically costly and hard to implement with sufficient accuracy.

\begin{figure}
	\centering
	\sidesubfloat[]{
		\includegraphics[scale=1]{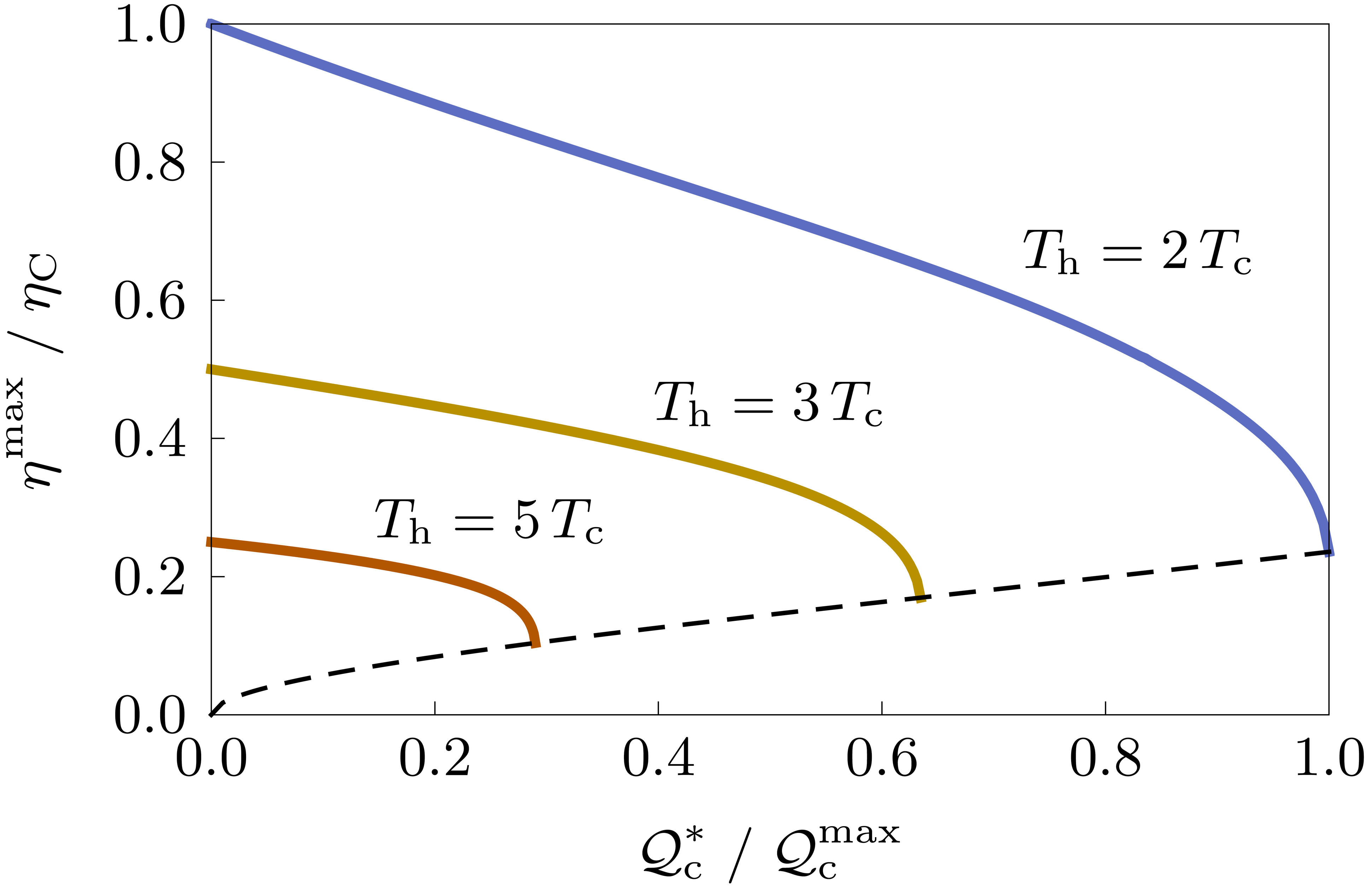}
		\label{fig:sc:cop_bound}
	}\\[1em]
	\sidesubfloat[]{
		\includegraphics[scale=1]{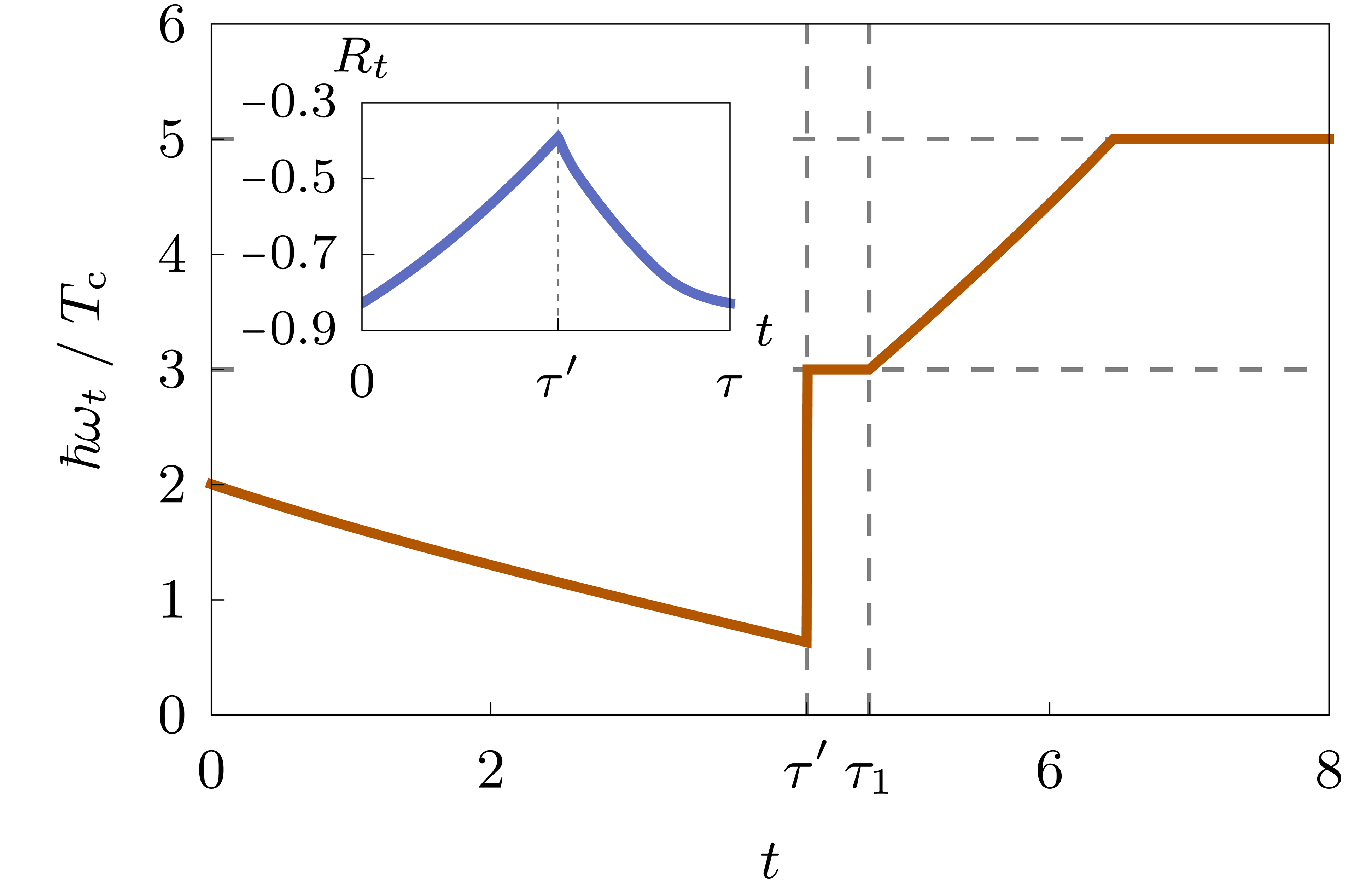}
		\label{fig:sc:cop_protocol}
	}
	\caption[necessary]{Microcooler at optimal efficiency.
		\begin{enumerate*}[(a)]
		\item Maximum efficiency as a function of the given heat extraction $\mathcal Q_\txtc^\ast$.
			The horizontal axis has been rescaled with the maximum heat extraction for $T_\txth = 2\, T_\txtc$, $\mathcal Q_\txtc^\txtmax$, and the vertical axis with the Carnot efficiency $\eta_C$ for the same temperatures.
			The dashed line shows how the efficiency at maximal cooling power decays as the temperature gradient becomes larger.
		\item Optimal control protocol leading to maximal efficiency for fixed heat extraction $\mathcal Q_\txtc^\ast = 0.9\, \mathcal Q_\txtc^\txtmax$.
			The inset shows the corresponding trajectory of the state variable $R_t$.
		\end{enumerate*}
		Throughout this figure, we have set $\switch\omega = 3\, T_\txtc / \hbar$, $\omega_\txtmax = 5\, T_\txtc / \hbar$, $T_\txth = 2\, T_\txth$ and $\period = 8\, \gamma^{-1}$.
		For these parameter values, the maximum extracted heat at $T_\txth = 2\, T_\txtc$ is $\mathcal Q_\txtc^\txtmax \approx 0.297\, T_\txtc$.
	}
	\label{fig:sc:cop_full}
\end{figure}

In the following, we describe a more practical way of finding the optimal reset protocol.
To this end, we first note that the Hamiltonian \eqref{eq:sc:efficiency_Hr} is, up to its sign, identical with \eqref{eq:sc:output_Hw}.
Thus, if $\Xreset H$ admits a local minimum with respect to $\omega_t$ in the range $[\switch\omega, \omega_\txtmax]$, the canonical equations can be solved exactly and the reset protocol reads
\begin{equation} \label{eq:sc:solution_cop}
	\omega_t = \frac{T_\txtc}{\hbar} \log\left[ \frac{2 - 2 C_1\, W_0[C_2 e^{-\gamma t}]}{C_1\, W_0[C_2 e^{-\gamma t}]^2} - 1 \right] ,
\end{equation}
where $C_1$ and $C_2$ are constants.
Note that, in contrast to \eqref{eq:sc:solution_w}, this solution must involve the upper rather than the lower branch of the Lambert $W$ function to ensure that the state variable decreases during the reset, i.e., $\dot R_t = F[R_t, \omega_t] < 0$.
According to Pontryagin's principle, the protocol $\effX\omega_t$ either follows the monotonically increasing trajectory \eqref{eq:sc:solution_cop} or takes on one of the boundary values $\switch\omega$ or $\omega_\txtmax$.
Consequently, if we assume that the optimal protocol does not jump within the reset stroke, it must have the general form shown in Fig.~\ref{fig:sc:cop_protocol}.
Specifically, $\effX\omega_t$ must be constant at $\switch\omega$ until a certain time $\period_\depart$, then follow \eqref{eq:sc:solution_cop} until it reaches $\omega_\txtmax$, and finally remain constant until the end of the stroke.
Since each protocol of this type is uniquely determined by the departure time $\period_\depart$, this procedure induces a one-to-one mapping between $\period_\depart$ and the state of the system at the end of the reset stroke, $R_\period = R_\period[\period_\depart]$.
This map can be determined analytically from the corresponding Bloch equation.
The only numerical operation that is required to determine the optimal reset protocol thus consists in solving the condition $R_\period[\period_\depart] = R_0$ for $\period_\depart$ \footnote{
		Note that, in order to account for reset protocols with $\omega_{\switch\period} > \switch\omega$, $\period_\depart$ must be allowed to become smaller than $\switch\period$.
	}.

The method described above makes it possible to find the protocol $\effX\omega_t[R_0, \omega_0, \mathcal Q_\txtc^\ast]$ that maximizes the efficiency of the cooling cycle for given $R_0$, $\omega_0$ and $\mathcal Q_\txtc^\ast$.
Inserting this protocol into \eqref{eq:fridge:efficiency} and optimizing the resulting function $\eta[R_0, \omega_0, \mathcal Q_\txtc^\ast]$ with respect to the initial values $R_0$ and $\omega_0$ finally yields the maximal efficiency at given cooling power.

This figure of merit is plotted in Fig.~\ref{fig:sc:cop_full} together with the corresponding optimal protocol; it approaches the Carnot limit \eqref{eq:carnot} for $\mathcal Q_\txtc^\ast \to 0$ and monotonically decays as $\mathcal Q_\txtc^\ast$ becomes larger.
Thus, increasing the heat extraction of the microcooler inevitably reduces its maximal efficiency.
This result aligns well with recent discoveries of universal trade-off relations between the extracted heat and the efficiency of mesoscopic thermal devices \cite{BrandnerPhysRevX2015,BauerPhysRevE2016,BrandnerPhysRevE2016,ShiraishiPhysRevLett2016,PietzonkaPhysRevLett2018,ShiraishiJStatPhys2019}.
Furthermore, Fig.~\ref{fig:sc:cop_bound} shows that not only the maximal cooling power but also the overall efficiency decays as the temperature of the hot reservoir becomes larger.
Hence, increasing the temperature bias is generally detrimental to the performance of the microcooler.

\section{Approximation Methods} \label{sec:approx}

\subsection{Rationale}

Our two-stroke scheme makes it possible to systematically optimize realistic models for mesoscopic thermal machines, as we have shown in the previous section for a superconducting microcooler.
To explore the optimal performance of even more complex devices, it is often helpful to first focus on limiting regimes, where dynamical approximation methods can be used to simplify computational tasks.
In this section, we develop such schemes for the key limits of slow or fast driving.
We thereby further extend our general framework and prepare the stage to investigate the thermodynamic performance of mesoscopic refrigerators in the coherent regime.

\subsection{Adiabatic Response} \label{subsec:approx:adiabatic}

We consider a slowly operated two-stroke refrigerator by assuming $\period, \switch\period \gg 1/\gamma$, where $\gamma$ is the typical relaxation rate of the working system.
Except for short transient periods at the beginning of each stroke, the system then follows the instantaneous equilibrium state $\vec R_\eq[\omega_t]$, which is defined by the condition
\begin{equation}
	\vec F[\vec R_\eq[\omega_t], \omega_t] \equiv 0 .
\end{equation}
In particular, we have
\begin{equation} \label{eq:gen:adiabatic}
	\vec R_{\switch\period} \simeq \vec R_\eq[\Xwork\omega]
	\quad\text{and}\quad
	\vec R_\period \simeq \vec R_\eq[\Xreset\omega] ,
\end{equation}
where $\Xwork\omega$ and $\Xreset\omega$ are the values of the control parameter at the end of the work and the reset stroke, respectively.
We note that this approximation can be systematically refined by including finite-time corrections.
To this end, the time-evolution equation \eqref{eq:gen:evolution} has to be solved perturbatively by expanding the state vector $\vec R_t$ in powers of the adiabaticity parameter $\varepsilon \equiv 1 / (\gamma\tau)$ \cite{CavinaPhysRevLett2017}.
However, to keep our analysis as transparent and simple as possible, we here neglect contributions of order $\varepsilon$.
The relations \eqref{eq:gen:adiabatic} significantly reduce the interdependence of work and reset stroke, and thus simplify our optimization scheme as follows.

To maximize the heat extraction \eqref{eq:gen:output}, the work protocol has to be found by solving the canonical equations \eqref{eq:gen:canonical_work} for fixed initial conditions $\vec R_0$ and $\vec\lambda_0$.
Since $\vec R_t$ does not change during the quenches of $\omega_t$, we now have $\vec R_0 = \vec R_\eq[\Xreset\omega]$, i.e., the initial state of the system is determined by one parameter $\Xreset\omega$.
Moreover, to restore this state after the work stroke, it suffices to set $\omega_t = \Xreset\omega$ for a short time $\Xreset\durationFunctional \simeq 1/\gamma \equiv \varepsilon\tau$.
Hence, in the zeroth order with respect to $\varepsilon$, we have $\period \simeq \switch\period$ and the reset stroke does not have to be optimized separately.
In fact, the optimal protocol $\outputX\omega_t$ is obtained by extending the work stroke over the entire cycle time $\period$ and maximizing the resulting heat extraction over $N+1$ parameters given by $\vec\lambda_0$ and $\Xreset\omega$.

Our second optimization criterion requires us to minimize the dissipated heat \eqref{eq:gen:cost} for given cooling power $\fixedOutput / \period$.
To this end, both strokes have to be taken into account.
Specifically, after finding the optimal work protocol $\omega_t[\Xreset\omega, \vec\lambda_0]$ as before, we first have to determine the switching time $\switch\period[\Xreset\omega, \vec\lambda_0, \fixedOutput]$ such that $\mathcal Q_\txtc[\omega_t] = \fixedOutput$, cf.~\eqref{eq:gen:switching_time_eff}.
To find the optimal reset protocol, the objective functional \eqref{eq:gen:ext_cost} has to be minimized using fixed initial conditions $\vec R_{\switch\period} = \vec R_\eq[\Xwork\omega]$ and $\vec\lambda_{\switch\period}$ for the state variables and Lagrange multipliers, respectively.
Here, $\Xwork\omega$ is determined by $\Xreset\omega$ and $\vec\lambda_0$; $\vec\lambda_{\switch\period}$ has to be chosen such that the cycle condition $\omega_\tau = \Xreset\omega$ is satisfied.
Owing to this constraint, the optimal protocol $\effX\omega_t[\Xreset\omega, \vec\lambda_0, \vec\lambda_{\switch\period}, \fixedOutput]$ effectively depends on $2N$ free parameters, which have to be eliminated by minimizing the corresponding heat release $\mathcal Q_\txth[\Xreset\omega, \vec\lambda_0, \vec\lambda_{\switch\period}]$.
Though generally non-trivial, this procedure is still significantly simpler than the full optimization, which involves $N$ boundary conditions to ensure that $\vec R_\tau = \vec R_0$.
By contrast, here only one constraint has to be respected.
The continuity of the state $\vec R_t$ is then enforced by the adiabaticity condition \eqref{eq:gen:adiabatic}.

\subsection{High-Frequency Response} \label{subsec:approx:highF}

Having understood how to optimize a two-stroke refrigerator in adiabatic response, we now consider the opposite limit $\period, \switch\period \ll 1/\gamma$.
In this regime, the state vector $\vec R_t$ changes only slightly during the individual strokes, since the working system is unable to follow the rapid variations of the control parameter $\omega_t$.
Therefore, we can use the approximations
\begin{align} \label{eq:gen:highF}
	\vec R_t &\simeq \vec R_0 + t \dot{\vec R}_0 = \vec R_0 + t \vec F[\vec R_0, \omega_0] \quad\text{and} \\
	\vec R_t &\simeq \vec R_{\switch\period} + (t - \switch\period) \dot{\vec R}_{\switch\period} = \vec R_{\switch\period} + (t - \switch\period) \vec F[\vec R_{\switch\period}, \omega_{\switch\period}] \nonumber
\end{align}
to describe the work and the reset stroke, respectively.
The initial states $\vec R_0$ and $\vec R_{\switch\period}$ are thereby fully determined as functions of $\omega_0$ and $\omega_{\switch\period}$ by the requirement that $\vec R_t$ is continuous throughout the cycle.
Thus, inserting the expansions \eqref{eq:gen:highF} into \eqref{eq:gen:output} and \eqref{eq:gen:cost} and neglecting second-order corrections in $1/\varepsilon \equiv \gamma\tau$ yields
\begin{align} \label{eq:gen:highF_output}
	\mathcal Q_\txtc[\omega_t] &\simeq \switch\period Q[\vec R_0, \omega_0] \equiv \mathcal Q_\txtc[\switch\period, \omega_0, \omega_{\switch\period}] \quad\text{and} \\
	\mathcal Q_\txth[\omega_t] &\simeq (\switch\period - \period) Q[\vec R_{\switch\period}, \omega_{\switch\period}] \equiv \mathcal Q_\txth[\switch\period, \omega_0, \omega_{\switch\period}] . \nonumber
\end{align}

These expressions show that both the extracted and the released heat of the device now depend only on the switching time $\switch\period$ and the initial values of the work and the reset protocols, $\omega_0$ and $\omega_{\switch\period}$.
Consequently, any control protocol $\omega_t$ can be mimicked with a step profile
\begin{equation} \label{eq:gen:highF_constant}
	\omega^{\text{HF}}_t[\switch\period, \omega_0, \omega_{\switch\period}] = \omega_0 + (\omega_{\switch\period} - \omega_0) \Theta[t - \switch\period] ,
\end{equation}
where $\Theta$ denotes the Heaviside function.
In particular, the optimal protocols $\outputX\omega_t$ and $\effX\omega_t[\fixedOutput]$ adopt the form \eqref{eq:gen:highF_constant} in the fast-driving limit.
For $\outputX\omega_t$, the free parameters $\switch\period$, $\omega_0$ and $\omega_{\switch\period}$ must be determined by maximizing $\mathcal Q_\txtc[\switch\period, \omega_0, \omega_{\switch\period}]$.
Analogously, $\effX\omega_t[\fixedOutput]$ is found by minimizing $\mathcal Q_\txth[\switch\period, \omega_0, \omega_{\switch\period}]$ under the constraint $\mathcal Q_\txtc[\switch\period, \omega_0, \omega_{\switch\period}] = \fixedOutput$.

The high-frequency approximation provides a simple yet powerful tool to explore the performance limits of mesoscopic refrigerators.
In fact, due to the universal form \eqref{eq:gen:highF_constant} of the high-frequency protocol, our general scheme can be reduced to relatively simple 3-parameter optimizations.
Moreover, the approximations \eqref{eq:gen:highF} and \eqref{eq:gen:highF_output} can be systematically refined by including higher-order corrections in $1/\varepsilon$, and thus introducing more and more variational parameters given by the higher derivatives of $\omega_t$ at $t=0$ and $t=\switch\period$.

\subsection{Semiclassical Microcooler Revisited} \label{subsec:approx:sc}

\begin{figure}
	\centering
	\includegraphics[scale=1]{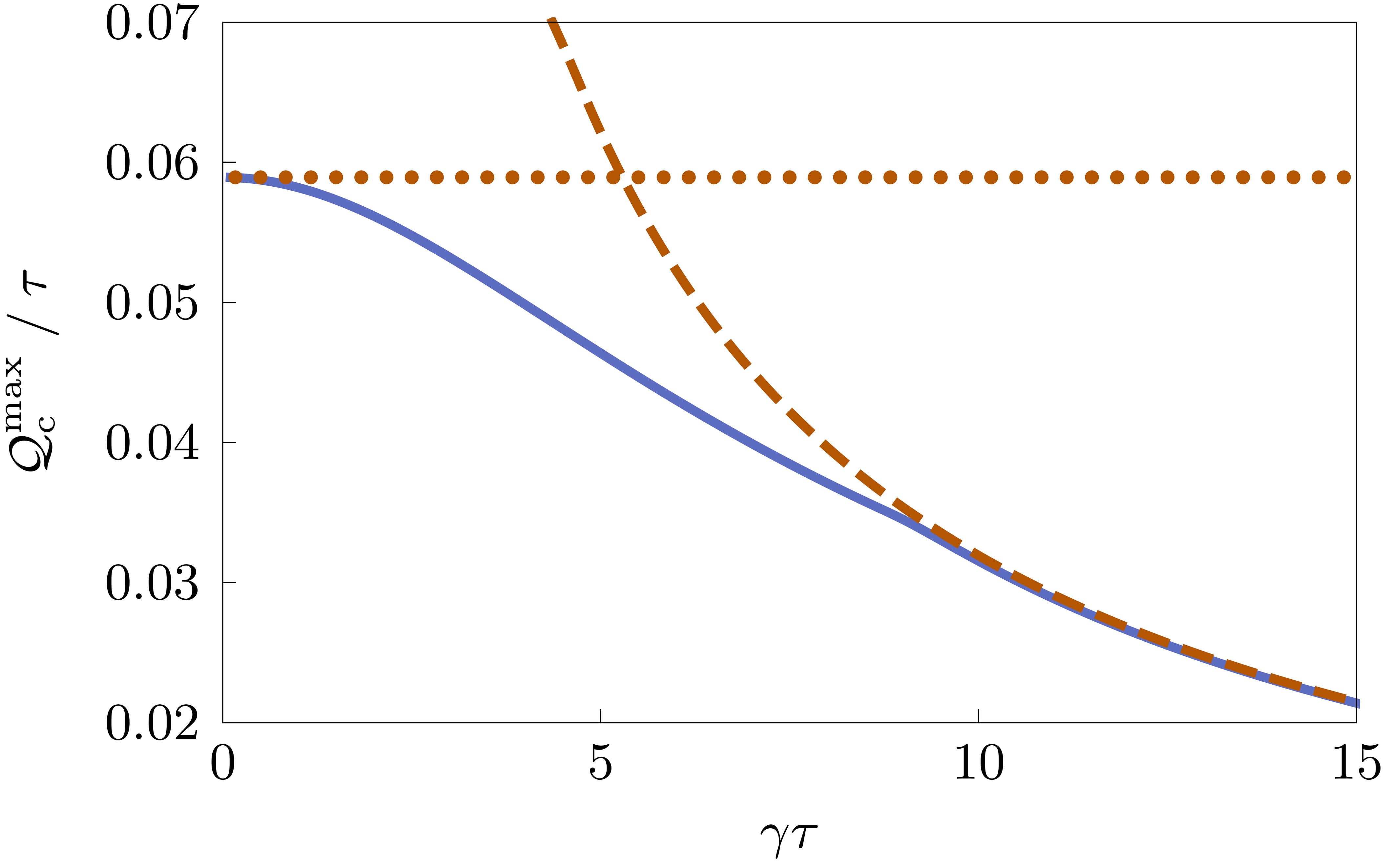}
	\caption{Quantum microcooler at slow and fast driving.
		The plot shows the maximum cooling power $\mathcal Q_\txtc^\txtmax / \period$ in units of $\gamma T_\txtc$ as a function of the inverse adiabaticity parameter $1/\varepsilon = \gamma\period$.
		In the limits $\gamma\period \gg 1$ and $\gamma\period \ll 1$, the exact result from Sec.~\ref{subsec:sc:cooling_power} (solid line) approaches the adiabatic (dashed line) and high-frequency (dotted line) approximations, respectively.
		The parameters in this figure are the same as in Fig.~\ref{fig:sc:results}, i.e., the blue curves in both plots are identical.}
	\label{fig:sc:approximations}
\end{figure}

Before moving on to the full quantum regime, we now illustrate our approximation scheme for the semiclassical microcooler.
For the sake of brevity, we here focus on maximum cooling power as our optimization criterion.

In the adiabatic limit, the reset stroke does not have to be considered explicitly and the optimal protocol $\outputX\omega_t$ is given by \eqref{eq:sc:solution_w} for $0 \leq t \leq \period$.
The two constants $C_1$ and $C_2$ thereby have to be chosen such that the extracted heat $\mathcal Q_\txtc[\omega_t] = \mathcal Q_\txtc[C_1, C_2]$ becomes maximal.
This condition is equivalent to optimizing the reset level $\Xreset\omega \in [\switch\omega, \omega_\txtmax]$ of the control parameter and the initial value $\lambda_0$ of the Lagrange multiplier, as described in the first part of Sec.~\ref{subsec:approx:adiabatic}.

The resulting optimal cooling power is shown in Fig.~\ref{fig:sc:approximations} as a function of the dimensionless cycle time $\gamma\period$, which corresponds to the inverse adiabaticity parameter $1/\varepsilon$.
This plot confirms that our adiabatic response scheme is indeed accurate for $\varepsilon \ll 1$.
In fact, the adiabatic approximation for $\mathcal Q_\txtc^\txtmax / \period$ departs from the exact result obtained in Sec.~\ref{subsec:sc:cooling_power} only at $1 / \varepsilon \equiv \gamma\tau \simeq 10$.

In the fast driving regime, the cooling power is maximized by a step protocol with the general form \eqref{eq:gen:highF_constant}.
The variational parameters $\omega_0$, $\omega_{\switch\period}$ and $\switch\period$ can be determined exactly by maximizing the heat extraction \eqref{eq:sc:cooling_power} after inserting \eqref{eq:gen:highF} and \eqref{eq:gen:highF_constant} and neglecting second order corrections in $1/\varepsilon = \gamma\period$.
We find that the optimal switching time is given by
\begin{equation}
	\period^\ast[\omega_0, \omega_{\switch\period}] \equiv \frac{\sqrt{\Gamma^+_0 \Gamma^+_{\switch\period}} - \Gamma^+_{\switch\period}}{\Gamma^+_0 - \Gamma^+_{\switch\period}}\, \period
\end{equation}
as a function of the levels $\omega_0$ and $\omega_{\switch\period}$ of the protocol $\omega^{\text{HF}}_t$.
Here, we have used the abbreviation $\Gamma^+_t \equiv \Gamma^+[\omega_t]$.
The optimal reset level is $\omega_{\switch\period}^\ast = \omega_\txtmax$ and the optimal work level follows from maximizing the cooling power
\begin{equation} \label{eq:sc:highF_result}
	\mathcal Q_\txtc^\txtmax / \period = \max_{\omega_0}\, \bigl\{ \hbar\omega_0 \gamma \left( 1 - 2\, \period^\ast[\omega_0, \omega^\ast_{\switch\period}] / \period \right) \bigr\} .
\end{equation}
Note that this expression is independent of $\varepsilon$, since here we consider only the lowest order of the high-frequency expansion.
Still, as shown in Fig.~\ref{fig:sc:approximations}, the exact optimal cooling power approaches the constant value \eqref{eq:sc:highF_result} for $1/\varepsilon \lesssim 1$, thus confirming the validity of our approximation scheme for the fast-driving regime.

\section{Quantum Microcooler II -- Coherent Regime} \label{sec:coh}

As a key application of our approximation methods, we will now show how the cooling power of the microcooler illustrated in Fig.~\ref{fig:fridge} can be optimized in the full quantum regime.
To this end, we first recall the qubit Hamiltonian \eqref{eq:fridge:hamiltonian},
\begin{equation} \label{eq:qu:hamiltonian_again}
	H_t \equiv \frac{\hbar\Delta}{2} \sigma_x + \frac{\hbar\omega_t}{2} \sigma_z ,
\end{equation}
which describes the working system of this device.
If the tunneling energy $\Delta$ is not negligible, the periodic state that emerges due to cyclic variation of the control parameter $\omega_t$ features coherences between the two energy levels of the qubit.
The corresponding density matrix must therefore be parametrized in the general form
\begin{equation} \label{eq:qu:parametrization}
	\rho_t \equiv \frac 1 2 ( \mathbbm 1 + \vec R_t\!\cdot\!\vec \sigma ) ,
\end{equation}
where $\vec\sigma \equiv (\sigma_x, \sigma_y, \sigma_z)^\intercal$ is the vector of Pauli matrices and the state vector $\vec R_t$ fulfills the Bloch equation \cite{NiskanenPhysRevB2007,KarimiPhysRevB2016}
\begin{align} \label{eq:qu:bloch}
	\dot{\vec R}_t
		&= \vec F[\vec R_t, \Omega_t] \\
		&\equiv \begin{bmatrix}
			-\Gamma^+_t \frac{\Omega_t^2 + \Delta^2}{2\Omega_t^2} & -\omega_t & -\Gamma^+_t \frac{\omega_t \Delta}{2\Omega_t^2} \\
			\omega_t & \!-\frac 1 2 \Gamma^+_t\! & -\Delta \\
			-\Gamma^+_t \frac{\omega_t \Delta}{2\Omega_t^2} & \Delta & -\Gamma^+_t \frac{2\Omega_t^2 - \Delta^2}{2\Omega_t^2}
		\end{bmatrix}\!\vec R_t - \frac{\Gamma^-_t}{\Omega_t} \begin{bmatrix} \Delta \\ 0 \\ \omega_t \end{bmatrix} . \nonumber
\end{align}
Here, the rates $\Gamma^\pm_t \equiv \Gamma^\pm[\Omega_t]$ are defined as in \eqref{eq:sc:bloch} with $\omega_t$ replaced by the instantaneous level splitting
\begin{equation}
	\Omega_t \equiv \sqrt{\Delta^2 + \omega_t^2} ,
\end{equation}
which we will treat as the effective control parameter of the system from here onwards.

In order to extend our step-rate model to the coherent regime, we have to take into account that $\Omega_t$ cannot vanish for finite $\Delta$.
Therefore, the lower bound $0$ in the coupling factor \eqref{eq:fridge:gamma} has to be replaced with $\Omega_\txtmin = \Delta$.
Furthermore, also the threshold frequency $\switch\Omega$, which now takes the role of $\switch\omega$ in the switching condition \eqref{eq:fridge:temp} for the reservoir temperature, has to be larger than $\Delta$.

Upon inserting \eqref{eq:qu:hamiltonian_again} and \eqref{eq:qu:parametrization} into the weak-coupling expression \eqref{eq:A} for the instantaneous heat flux, the mean heat extraction in the coherent regime becomes a functional of $\Omega_t$,
\begin{align} \label{eq:qu:obj_functional}
	\mathcal Q_\txtc[\Omega_t] &\equiv \int_0^{\switch\period} Q[\vec R_t, \Omega_t] \id t \\
		&\equiv -\int_0^{\switch\period} \!\left( \frac{\hbar\Gamma^+_t}{2} \bigl( \Delta R^x_t + \omega_t R^z_t \bigr) + \frac{\hbar\Gamma^-_t}{2} \Omega_t \right) \id t, \nonumber
\end{align}
which could, in principle, be optimized by applying the 3-step procedure of Sec.~\ref{subsec:general:output}.
This endeavor can be expected to be technically quite involved, since the periodicity constraint $\vec R_\period = \vec R_0$ now leads to three independent boundary conditions for the reset stroke, while only a single parameter is available to control the time-evolution of the state $\vec R_t$.
However, to understand how the optimal performance of the microcooler changes in the quantum regime, it is sufficient to determine the impact of the tunneling energy $\Delta$ on its maximum cooling power.
For this purpose, it is not necessary to carry out the full optimization procedure.
Instead, we can focus our analysis on the limits of slow and fast driving, where our approximation schemes enable a simple and physically transparent approach.

In the adiabatic-response regime, only the work stroke needs to be optimized \footnote{
		We note that for the adiabatic approximation to apply here, the relaxation time $1/\gamma$ has to be large compared to both the cycle time $\period$ and the time scale of the coherent oscillations.
	}.
To this end, we first integrate the canonical equations corresponding to the effective Hamiltonian
\begin{equation} \label{eq:qu:hamiltonian}
	\Xwork H[\vec R_t, \vec\lambda_t, \Omega_t] = Q[\vec R_t, \Omega_t] + \vec\lambda_t\!\cdot\!\vec F[\vec R_t, \Omega_t]
\end{equation}
for the given initial conditions
\begin{equation}
	\vec R_0 = \vec R_\eq[\Xreset\Omega] = -\frac{\Gamma^-[\Xreset\Omega]}{\Gamma^+[\Xreset\Omega]}\, \biggl( \frac{\Delta}{\Xreset\Omega}, 0, \frac{\sqrt{\Xreset\Omega^2 - \Delta^2}}{\Xreset\Omega} \biggr)^\intercal
\end{equation}
and $\vec\lambda_0$, see Appendix~\ref{sec:appendix2}.
The parameters $\Xreset\Omega$ and $\vec\lambda_0$ then have to be determined by maximizing the heat extraction $\mathcal Q_\txtc[\Omega_t] = \mathcal Q_\txtc[\Xreset\Omega, \vec\lambda_0]$.
This task is \emph{a priori} challenging, since the initial Lagrange multipliers $\lambda^x_0$ and $\lambda^y_0$ are left unbounded by physical constraints.
To overcome this problem, we use an iterative algorithm, which tracks the maximum of $\mathcal Q_\txtc[\Xreset\Omega, \vec\lambda_0]$ as $\Delta$ is increased in small steps starting from its semiclassical value $\Delta = 0$.
This approach relies on the implicit assumption that the global maximum of the function $\mathcal Q_\txtc[\Xreset\Omega, \vec\lambda_0]$ follows a continuous trajectory in the $4$-dimensional space of variational parameters, which is justified \emph{a posteriori} by the physical consistency of our results.

In the high-frequency regime, the cooling power is maximized by the step protocol
\begin{equation}
	\Omega^{\text{HF}}_t[\switch\period, \Omega_0, \Omega_{\switch\period}] = \Omega_0 + (\Omega_{\switch\period} - \Omega_0) \Theta[t - \switch\period] .
\end{equation}
As in the semiclassical case discussed in Sec.~\ref{subsec:approx:sc}, the variational parameters $\switch\period$, $\Omega_0$ and $\Omega_{\switch\period}$ can be determined by maximizing the corresponding cooling power in first order with respect to $1/\varepsilon = \gamma\period$.
The resulting expression for $\mathcal Q_\txtc[\switch\period, \Omega_0, \Omega_{\switch\period}]$ is rather involved and we do not show it here.
The optimal variational parameters can however be determined numerically.
We note in particular that this optimization yields $\Omega_{\switch\period}^\ast = \Omega_\txtmax$.

\begin{figure}
	\centering
	\sidesubfloat[]{
		\includegraphics[height=15em]{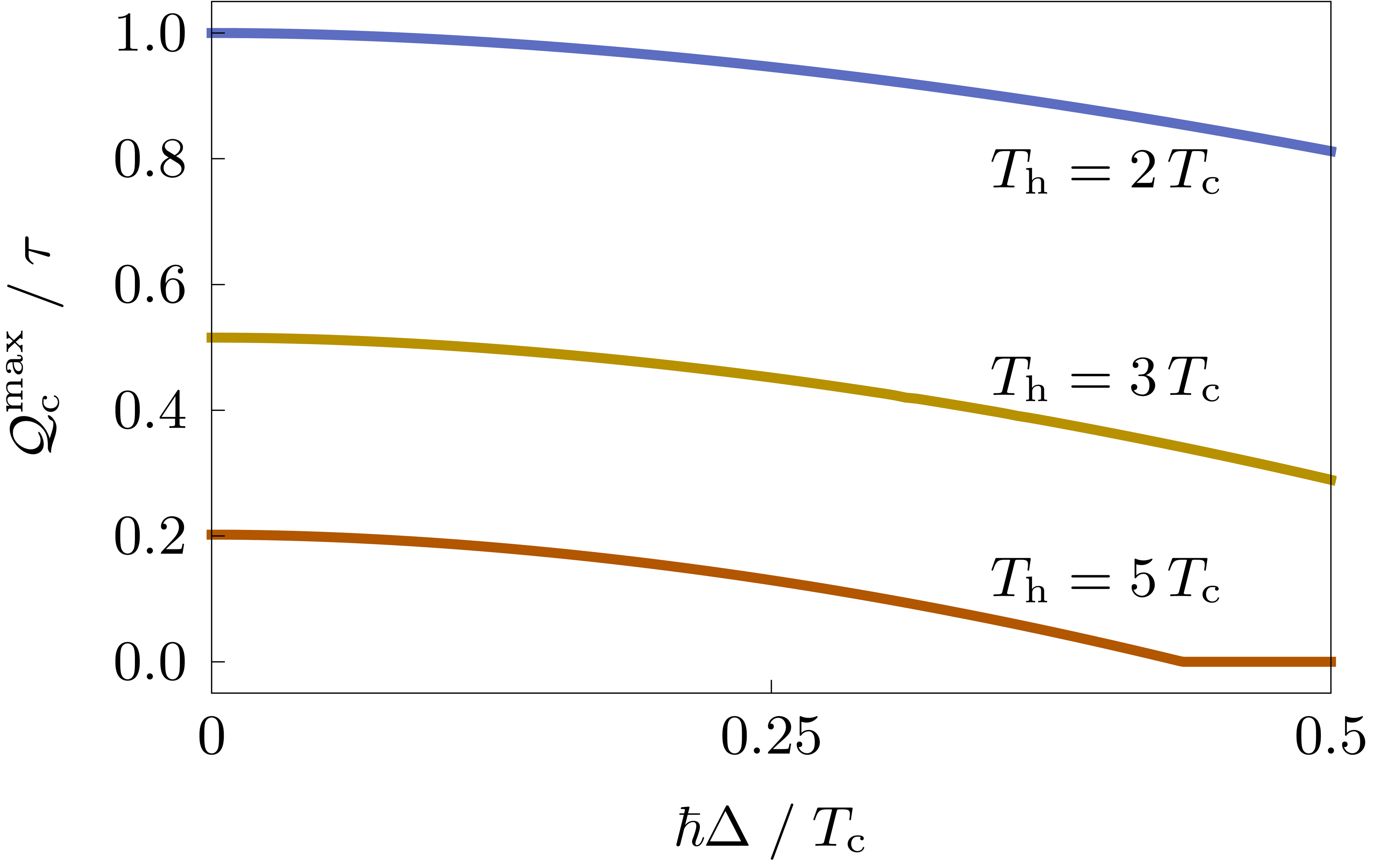}
		\label{fig:qu:adiabatic_results}
	}\\[1em]
	\sidesubfloat[]{
		\includegraphics[height=15em]{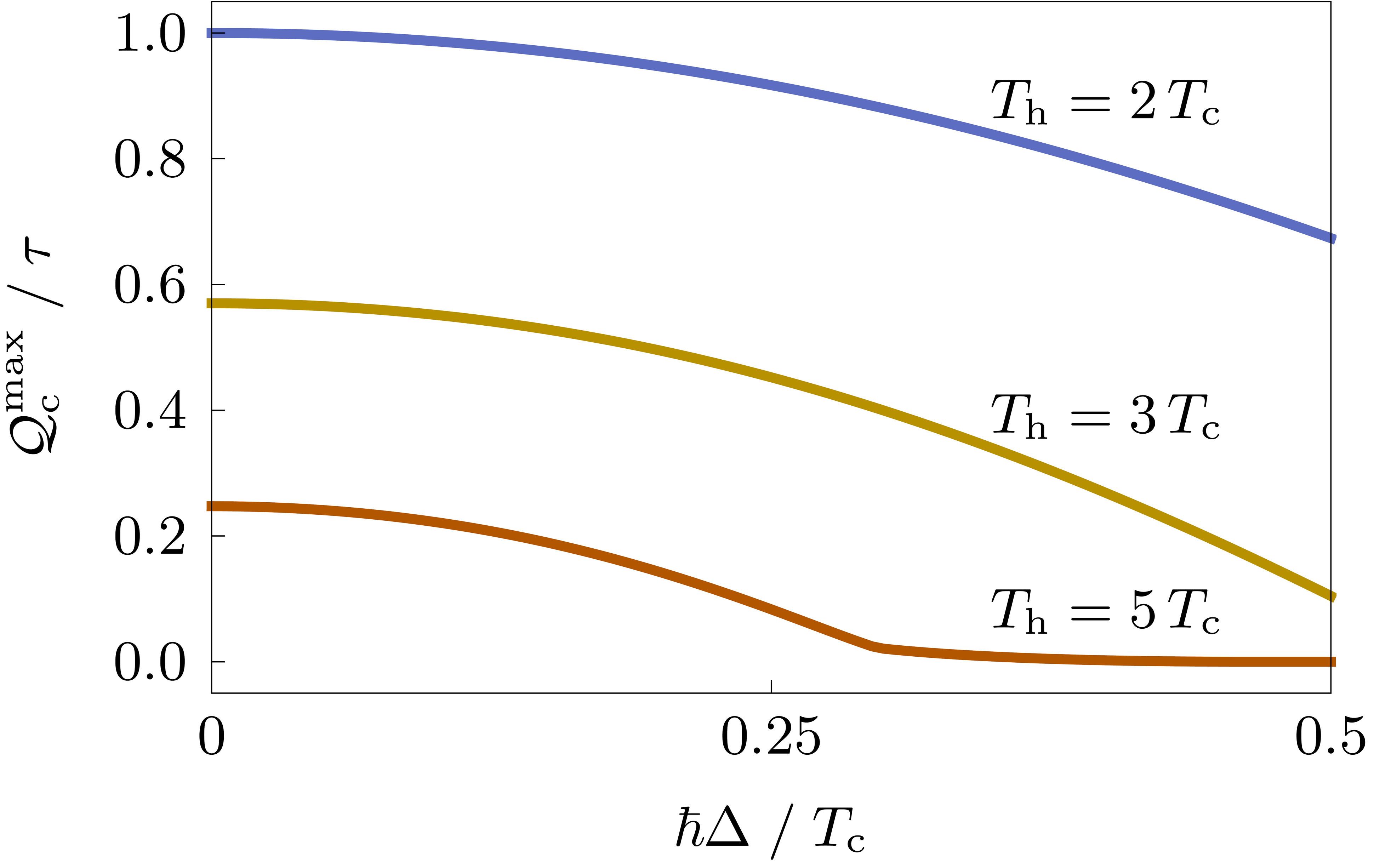}
		\label{fig:qu:highF_results}
	}
	\caption[necessary]{Optimal cooling power of the coherent microcooler as a function of the tunneling energy $\Delta$.
		\begin{enumerate*}[(a)]
		\item Maximum cooling power in the adiabatic regime ($\gamma\period = 10$) for different temperatures of the hot reservoir.
		\item Maximum cooling power in the high-frequency regime (independent of $\gamma\period$).
		\end{enumerate*}
		Here, we have used $\switch\Omega = 1.5\, T_\txtc / \hbar$, $\Omega_\txtmax = 3\, T_\txtc / \hbar$ and $\hbar\gamma = T_\txtc$.
		For comparison with the semiclassical model, the cooling power has been normalized with its value at $\Delta=0$ and $T_\txth = 2\, T_\txtc$ in both plots.
	}
	\label{fig:qu:results}
\end{figure}

Figure~\ref{fig:qu:results} shows the result of our analysis.
In both the adiabatic and the high-frequency limit, the maximum cooling power monotonically decreases from its semiclassical value to $0$ as $\Delta$ increases.
This behavior can be explained as follows.
The tunneling energy $\Delta$ corresponds to the minimal gap between the energy levels of the working system, see Fig.~\ref{fig:fridge}.
Increasing this parameter reduces the amount of thermal energy that can be absorbed during the work stroke.
As $\Delta$ approaches a certain critical value, the capacity of the working system to pick up heat from the cold reservoir becomes too small for the device to operate properly.
The optimal protocol then keeps the system practically in equilibrium at the low temperature $T_\txtc$ throughout the work stroke and the cooling power becomes zero.
Since this general picture can be expected to prevail also for intermediate driving speed, we can conclude that to engineer a powerful microcooler, the tunneling energy of the qubit must be kept as small as possible.

\section{Discussion and Outlook} \label{sec:outlook}

Our work provides a systematic scheme to optimize periodic driving protocols for mesoscopic two-stroke machines. 
Though developed here specifically for refrigerators, this general framework can easily be adapted to other types of thermal devices. 
Reciprocating heat engines, for example, use a periodically driven working system to convert thermal energy into mechanical power \cite{QuanPhysRevE2007,KosloffEntropy2013,VinjanampathyContempPhys2016,BenentiPhysRep2017}.
Within our two-stroke approach, this process can be described as a reversed cooling cycle. 
That is, the system picks up heat from a hot reservoir in the first stroke and returns to its initial state while being in contact with a cold reservoir in the second stroke.
To achieve optimal performance, the engine has to generate as much work output as possible from a given amount of thermal input energy.
Owing to the first law, this optimization criterion is equivalent to minimizing the dissipated heat during the reset while keeping the heat uptake during the work stroke fixed.
The corresponding optimal control protocol can thus be determined using the 3-step procedure of Sec.~\ref{subsec:general:efficiency}.

The performance figures of mesoscopic thermal devices, such as power and efficiency, can generally not be optimized simultaneously.  
Instead, they are subject to universal trade-off relations as several recent studies have shown \cite{Sothmann_2014,ShiraishiPhysRevLett2016,PietzonkaPhysRevLett2018,BrandnerPhysRevLett2017}.
As one of its potential key applications, our two-stroke scheme makes it possible to test the quality of these constraints under practical conditions.
Furthermore, covering both classical and quantum systems, the framework developed in this article might open a new avenue to systematically explore the impact of coherence on the performance of thermodynamic cycles, a central topic in quantum thermodynamics, see for example Refs.~\cite{BrunnerPhysRevE2014,UzdinPhysRevX2015,VinjanampathyContempPhys2016,RossnagelPhysRevLett2014,LostaglioPhysRevX2015,CwiklinskiPhysRevLett2015,BrandnerPhysRevLett2017,KlaersPhysRevX2017}.

To facilitate future investigations in these directions, our scheme can be combined with a variety of dynamical approximation methods.
In Sec.~\ref{sec:approx}, for example, we have shown how adiabatic and high-frequency expansion techniques can be included.
To this end, we have solved the dynamical constraint perturbatively assuming that the external driving is either slow or fast compared to the relaxation time of the working system.
This approach makes it possible to circumvent the use of Lagrange multipliers and thus reduces the amount of dynamical parameters in the optimization problem.
An alternative strategy could use the variational equations in the extended parameter space as a starting point.
Specifically, the canonical structure of these equations makes it possible to implement a variety of tools that were originally developed for the description of classical Hamiltonian systems including adiabatic gauge potentials \cite{KolodrubetzPhysRep2017}, shortcuts to adiabaticity \cite{JarzynskiPhysRevA2013,DeffnerPhysRevX2014} or non-linear generalizations of the Magnus expansion \cite{BlanesPhysRep2009}.

Integrating such advanced methods into our general framework will inevitably require a reliable reference to assess their practicality and accuracy. 
Such a testbed is provided in Sec.~\ref{sec:fridge}, where we have developed a simple and physically transparent model of a quantum microcooler, whose optimal operation cycle can be determined exactly.
In fact, this case study provides both a demonstration that our theoretical framework is directly applicable to ongoing experiments with engineered quantum systems and a valuable benchmark for further advances in theoretical optimization methods.

\begin{acknowledgments}
We thank J.~P.~Pekola for useful discussions.
K.~B. acknowledges support from the Academy of Finland (Contract No.~296073).
This work was supported by the Academy of Finland (projects No.~308515 and 312299).
All authors are associated with the Centre for Quantum Engineering at Aalto University.
\end{acknowledgments}

\appendix

\section{Alternative Optimization Scheme for the Semiclassical Microcooler} \label{sec:appendix1}

In Sec.~\ref{subsec:sc:cooling_power}, we have derived the optimal work protocol \eqref{eq:sc:solution_cop} for the semiclassical microcooler by enforcing the dynamical constraint \eqref{eq:sc:bloch} with a Lagrange multiplier.
Here, we present an alternative method to obtain the result \eqref{eq:sc:solution_cop}, which exploits the one-to-one correspondence between the control parameter and the derivative of the state variable in this model.

We proceed as follows.
First, solving Eq.~\eqref{eq:sc:bloch} for $\omega_t$ yields
\begin{equation} \label{eq:sc:level_splitting}
	\omega_t = \frac{T_\txtc}{\hbar} \log\left[ \frac{\gamma (1 - R_t)}{\dot R_t + \gamma (1 + R_t)} \right] .
\end{equation}
Upon inserting this expression into \eqref{eq:sc:cooling_power}, the objective functional becomes
\begin{equation} \label{eq:sc:Qc_appendix}
	\mathcal Q_\txtc[R_t] = \int_0^{\switch\period} Q[R_t, \dot R_t] \id t ,
\end{equation}
where the effective Lagrangian
\begin{equation}
	Q[R_t, \dot R_t] \equiv \frac{T_\txtc}{2} \dot R_t\, \log\left[ \frac{\gamma (1 - R_t)}{\dot R_t + \gamma (1 + R_t)} \right]
\end{equation}
does not explicitly depend on time.
Consequently, the corresponding effective Hamiltonian is a constant of motion given by
\begin{equation} \label{eq:sc:ode}
	4\gamma\, C_1 \equiv \frac{\dot R_t^2}{\dot R_t + \gamma (1 + R_t)} .
\end{equation}
Using \eqref{eq:sc:bloch}, $C_1$ can be expressed in terms of the initial values $R_0$ and $\omega_0$ as
\begin{equation} \label{eq:sc:c1}
	C_1 = \frac{1}{1 - R_0} \left( R_0 \cosh\left[ \frac{\hbar\omega_0}{2 T_\txtc} \right] + \sinh\left[ \frac{\hbar\omega_0}{2 T_\txtc} \right] \right)^2 .
\end{equation}
This expression shows that $C_1$ is non-negative.
Furthermore, for $C_1 = 0$, \eqref{eq:sc:ode} and \eqref{eq:sc:c1} imply $R_t = R_0 = -\tanh[ \hbar\omega_0 / (2T_\txtc) ]$ and $\omega_t = \omega_0$, that is, the system is in equilibrium throughout the cycle and the average heat extraction \eqref{eq:sc:Qc_appendix} becomes zero.

Second, solving \eqref{eq:sc:ode} for $\dot R_t$ gives
\begin{equation} \label{eq:sc:ode2}
	\dot R_t = 2\gamma \big( C_1 \pm \sqrt{C_1 (C_1 + 1 + R_t)} \big) ,
\end{equation}
where only the positive branch of the square root leads to $\dot R_t > 0$ and thus positive heat extraction.
Since we require that the control parameter $\omega_t$, which is given by \eqref{eq:sc:level_splitting} in terms of $R_t$, does not jump during the work stroke, both $R_t$ and $\dot R_t$ must be continuous.
We can thus neglect the negative branch in \eqref{eq:sc:ode2}.
Note that this choice implies the constraint $\dot R_0 = F[R_0, \omega_0] > 0$ on the initial values $R_0$ and $\omega_0$, cf.~\eqref{eq:sc:ic_restriction} and \eqref{eq:sc:ic_restrictions_final}.

Third, solving the differential equation \eqref{eq:sc:ode2} under this condition yields
\begin{equation}
	\outputX R_t = -1 + C_1 \left( \left( 1 + W_{-1}[C_2 e^{-\gamma t}] \right)^2 - 1 \right) ,
\end{equation}
where the dimensionless constant $C_2$ is given by
\begin{equation}
	C_2 = W^{-1}\!\left[ \frac{2 (1 - R_0)}{(1 + R_0)\, e^{\hbar\omega_0 / (kT_\txtc)} - (1 - R_0)} \right] ,
\end{equation}
with $W^{-1}[x] \equiv x\e^x$.
Thus, we have recovered the result \eqref{eq:sc:solution_R} of the main text.

\section{Optimal Work Stroke of the Coherent Microcooler} \label{sec:appendix2}

The optimal work stroke of the coherent microcooler discussed in Sec.~\ref{sec:coh} is described by the effective Hamiltonian
\begin{align} \label{eq:qu:Hw}
	&\Xwork H[\vec R_t, \vec\lambda_t, \Omega_t] \\
	&= \frac{T_\txtc}2 \vec\lambda_t\!\cdot\!\vec F[\vec R_t, \Omega_t] - \frac{\hbar\Gamma^+_t}{2} \bigl( \Delta R^x_t + \omega_t R^z_t \bigr) - \frac{\hbar\Gamma^-_t}{2} \Omega_t , \nonumber
\end{align}
where we rescaled the Lagrange multipliers $\vec\lambda_t$ by a factor of $T_\txtc / 2$ compared with \eqref{eq:qu:hamiltonian} for convenience.
The corresponding canonical equations for the state variables and Lagrange multipliers are given by
\begin{align} \label{eq:qu:bloch2}
	\dot{\vec R}_t &= \vec F[\vec R_t, \Omega_t] \\
		&\equiv \begin{bmatrix}
			-\Gamma^+_t \frac{\Omega_t^2 + \Delta^2}{2\Omega_t^2} & -\omega_t & -\Gamma^+_t \frac{\omega_t \Delta}{2\Omega_t^2} \\
			\omega_t & \!-\frac 1 2 \Gamma^+_t\! & -\Delta \\
			-\Gamma^+_t \frac{\omega_t \Delta}{2\Omega_t^2} & \Delta & -\Gamma^+_t \frac{2\Omega_t^2 - \Delta^2}{2\Omega_t^2}
		\end{bmatrix} \vec R_t - \frac{\Gamma^-_t}{\Omega_t} \begin{bmatrix} \Delta \\ 0 \\ \omega_t \end{bmatrix} \nonumber
\end{align}
and
\begin{equation} \label{eq:qu:cobloch}
	\dot{\vec\lambda}_t = \begin{bmatrix}
			\Gamma^+_t \frac{\Omega_t^2 + \Delta^2}{2\Omega_t^2} & -\omega_t & \Gamma^+_t \frac{\omega_t \Delta}{2\Omega_t^2} \\
			\omega_t & \frac 1 2 \Gamma^+_t & -\Delta \\
			\Gamma^+_t \frac{\omega_t \Delta}{2\Omega_t^2} & \Delta & \Gamma^+_t \frac{2\Omega_t^2 - \Delta^2}{2\Omega_t^2}
		\end{bmatrix}\vec\lambda_t + \frac{\hbar\Gamma^+_t}{T_\txtc} \begin{bmatrix} \Delta \\ 0 \\ \omega_t \end{bmatrix} ,
\end{equation}
respectively.
We recall that the energy bias $\omega_t$ and the level splitting $\Omega_t$ are related by $\omega_t = \sqrt{\Omega_t^2 - \Delta^2}$.

The evolution equations \eqref{eq:qu:bloch2} and \eqref{eq:qu:cobloch} are coupled by the algebraic constraint
\begin{equation} \label{eq:qu:constraint}
	\frac{\partial}{\partial\Omega_t} \Xwork H[\vec R_t, \vec\lambda_t, \Omega_t] = 0.
\end{equation}
This differential-algebraic system could, in principle, be integrated by solving the algebraic constraint for $\Omega_t = \Omega[\vec R_t, \vec\lambda_t]$.
Equations \eqref{eq:qu:bloch2} and \eqref{eq:qu:cobloch} would then become an ordinary system of differential equations, which could be integrated using standard techniques.
This approach has been used for the semiclassical microcooler in Sec.~\ref{subsec:sc:cooling_power}.
However, owing to its complicated structure, solving the constraint \eqref{eq:qu:constraint} for $\Omega_t$ is hard to implement in practice.

Instead, it is more convenient to transform the Bloch equations into a co-rotating frame.
To this end, we define the transformed Bloch vector $\vec r_t$ by replacing the static parametrization \eqref{eq:qu:parametrization} with
\begin{equation} \label{eq:qu:param2}
	\rho_t \equiv \frac 1 2 V_t (\mathbbm 1 + \vec r_t\!\cdot\!\vec \sigma) V_t^\dagger .
\end{equation}
Here, $V_t$ denotes the unitary matrix
\begin{equation}
	V_t \equiv \begin{pmatrix}
			\cos[\varphi_t/2] & -\sin[\varphi_t/2] \\
			\sin[\varphi_t/2] & \cos[\varphi_t/2]
		\end{pmatrix}
\end{equation}
with $\tan[\varphi_t] \equiv \Delta / \omega_t$, which diagonalizes the instantaneous Hamiltonian $H_t$.
In fact, the vectors $\vec R_t$ and $\vec r_t$ differ by a rotation in the $x$-$z$ plane by the angle $\varphi_t$.
This change of coordinates separates the population and the coherence degrees of freedom of the density matrix, which are now parametrized by $r^z_t$ and $r^{x,y}_t$, respectively.
Note that, in contrast to $\vec R_t$, the transformed Bloch vector $\vec r_t$ is not continuous at the jumps of the control protocol $\Omega_t$;
	if $V_{t-\dt}$ and $V_t$ are the rotation operators corresponding to the Hamiltonian before and after the jump, respectively, the accompanying jump in $\vec r_t$ is determined by the condition
\begin{equation}
	r^k_t = \tr\bigl[ V_t^\dagger V_{t-\dt}\, (\vec r_{t - \dt}\!\cdot\!\vec\sigma)\, V_{t-\dt}^\dagger V_t\; \sigma_k \bigr] / 2 .
\end{equation}

In the following, we will show how the optimal work protocol can be calculated in the rotating frame.
To this end, we first observe that the transformed Bloch equation reads
\begin{equation} \label{eq:qu:bloch_transf}
	\dot{\vec r}_t = \begin{bmatrix}
			-\Gamma^+_t / 2 & -\Omega_t & -\dot\varphi[\Omega_t, \dot\Omega_t] \\
			\Omega_t & -\Gamma^+_t / 2 & 0 \\
			\dot\varphi[\Omega_t, \dot\Omega_t] & 0 & -\Gamma^+_t
		\end{bmatrix} \vec r_t - \begin{bmatrix} 0 \\ 0 \\ \Gamma^-_t \end{bmatrix} .
\end{equation}
As an artifact of the time-dependent parametrization \eqref{eq:qu:param2}, the right hand side of \eqref{eq:qu:bloch_transf} now depends on the time-derivative of the control parameter, $\dot\Omega_t$.
Our general optimization scheme can, however, still be applied without major modifications since $\dot\Omega_t$ has no physical significance here.

The transformed vector of Lagrange multipliers, $\vec\Lambda_t$, satisfies the evolution equation
\begin{equation} \label{eq:app:cobloch_transf}
	\dot{\vec\Lambda_t} = \begin{bmatrix}
			\Gamma^+_t / 2 & -\Omega_t & -\dot\varphi[\Omega_t, \dot\Omega_t] \\
			\Omega_t & \Gamma^+_t / 2 & 0 \\
			\dot\varphi[\Omega_t, \dot\Omega_t] & 0 & \Gamma^+_t
		\end{bmatrix} \vec\Lambda_t + \frac{\hbar\Omega_t}{T_\txtc} \begin{bmatrix} 0 \\ 0 \\ \Gamma^+_t \end{bmatrix}
\end{equation}
in the rotating frame and the algebraic constraint \eqref{eq:qu:constraint} becomes
\begin{widetext}
\begin{align} \label{eq:app:algebraic_transf}
	&\frac{\hbar\Delta}{T_\txtc} \left\{ 2\Lambda^x_t \left( 1 - \e^{-W_t} \right) + \left( 2W_t\, r^x_t + \Lambda^z_t r^x_t + \Lambda^x_t r^z_t \right) \left( 1 + \e^{-W_t} \right) + \frac{2T_\txtc}{\hbar\gamma}\, W_t \left( \Lambda^y_t r^z_t - \Lambda^z_t r^y_t \right) \right\} \\
	&= \frac{\hbar\Omega_t}{T_\txtc} \frac{\hbar\omega_t}{T_\txtc} \left\{ 2 (1 + r^z_t) - \frac{2T_\txtc}{\hbar\gamma} \left( \Lambda^y_t r^x_t - \Lambda^x_t r^y_t \right) - \e^{-W_t} \Bigl( 2(1-r^z_t)(1-W_t-\Lambda^z_t) + \Lambda^x_t r^x_t + \Lambda^y_t r^y_t \Bigr) \right\} \nonumber
\end{align}
\end{widetext}
in the new variables, where $W_t = \hbar\Omega_t / T_\txtc$.

In order to obtain a closed system of differential equations, we have to express $\dot\Omega_t$ in terms of $\vec r_t$, $\vec\Lambda_t$ and $\Omega_t$.
To this end, we take the time-derivative of the algebraic constraint \eqref{eq:app:algebraic_transf} and then use \eqref{eq:qu:bloch_transf} and \eqref{eq:app:cobloch_transf} to eliminate $\dot{\vec r}_t$ or $\dot{\vec\Lambda}_t$.
The resulting expression can be rewritten in the form
\begin{equation} \label{eq:app:d_algebraic}
	\dot\Omega_t = \dot\Omega[\vec r_t, \vec\Lambda_t, \Omega_t] .
\end{equation}

The relation \eqref{eq:app:d_algebraic} enables the following strategy to find the optimal time evolution.
We first choose initial values $(\vec r_0, \vec\Lambda_0, \Omega_0)$, which are compatible with the algebraic constraint \eqref{eq:app:algebraic_transf}.
For this purpose, we note that \eqref{eq:app:algebraic_transf} is a linear equation in $\vec r_t$ and $\vec \Lambda_t$.
Therefore, it is straightforward to determine, for example, $\Lambda^z_0$ if all other initial values are given.
The equations \eqref{eq:qu:bloch_transf}, \eqref{eq:app:cobloch_transf} and \eqref{eq:app:d_algebraic} then form an autonomous system of seven first-order differential equations, which can be treated as a standard initial value problem.
By construction, the resulting solution complies with the algebraic constraint \eqref{eq:app:algebraic_transf} at any time $t \geq 0$.

\bibliography{references}

\end{document}